\documentclass[aps,prd,onecolumn,groupedaddress,showpacs,nofootinbib,amssymb]{revtex4}
\usepackage[dvips]{graphicx}
\usepackage{amssymb}
\usepackage{amsmath}
\usepackage{graphicx}
\usepackage{amsfonts}
\usepackage{bm}
\usepackage{latexsym,float}
\usepackage{epsfig}
\usepackage{epstopdf}
\usepackage{amssymb}
\usepackage{verbatim}
\usepackage{multirow}
\usepackage{color}

\begin{document}

\title{On energy-momentum tensor for gravitational waves in $f(R)$ gravity}



\author{Petr V. Tretyakov}
\email{tpv@theor.jinr.ru} \affiliation{Joint Institute for
Nuclear Research, Joliot-Curie 6, 141980 Dubna, Moscow region,
Russia}

\author{Alexander N. Petrov}
\email{alex.petrov55@gmail.com} \affiliation{Sternberg Astronomical Institute, Moscow M.V. Lomonosov State University, Moscow,
Russia}

\begin{abstract}
The classical Isaacson's procedure for describing back-reaction of the averaged energy-momentum for high frequency gravitational waves is generalized to the $f(R)$ gravity case. From the beginning it is assumed that an initial  background could be arbitrary one.
Then, we restrict the
background to be de Sitter, which is a novelty regarding the
study of a back-reaction in $f(R)$ gravity.
 Consideration of the de Sitter space as a background spacetime allows us to provide the averaging procedure completely. Using the results on the de Sitter space and generalizing the Isaacson procedure, we construct the averaged energy-momentum on an additionally curved (averaged) background.  Consistency
tests for de Sitter spacetime are performed both at
the background and perturbative regimes. Our results generalize
previous studies in which the authors consider the
flat (Minkowski) spacetime as the initial background.
\end{abstract}


\maketitle

\section{Introduction}

One of the most important modern problems which are considered in gravitational theory is the compatibility
of General Relativity (GR), first, with the Standard Model of particle physics, second, with
the current astrophysical and cosmological observations. The latter shows the accelerated expansion
of the universe at its early epoch as well as at the present time, attributing the
inflation phase and the dark energy presence, respectively. In addition, the observations of galaxies signal on the
presence of a so called dark matter --- a special substance, which manifests through gravitational effects only. To escape various exotic models of particles, many modified theories of gravity are applied for potential explanations for these phenomena. More or less full list of many modified theories can be find in the reviews \cite{CFPS1,NO1.1,NOO}.

In the present paper, we are interested in $f(R)$ theories \cite{SF1,FT1,NO1}, which are probably the most  popular modified gravities, especially, among higher-order gravities. The reasons of such a popularity are as follows. From the one hand, $f(R)$ actions and related equations are very simple with respect to other higher-order theories such as 1) theories with various invariants of the curvature tensor, like $f(R,R^{ik}R_{ik},R^{ikmn}R_{ikmn})$ and others, see for example, \cite{IM,MI}; 2) theories with Gauss-Bonnet term, like $f(R,G)$ and others, see, for example, \cite{LBM}; 3)   theories with derivatives of the curvature tensor, like a simple variant $f(R,\Box R,\ldots,\Box^i R)$, see, for example, \cite{Schmidt}. Of course, $f(R)$ gravity is easier to be handled,  than the theories listed above  and analogous ones. From the other hand, $f(R)$ theories are sufficiently general to have the possibility to describe cosmological, astrophysical
or astronomical scenarios. Indeed, one of possibilities is when $f(R)$ can be presented as a series expansion in a scalar curvature $R$ \cite{SF1}:
\begin{equation}
f(R) =  \ldots + \frac{\alpha_2}{R^2} + \frac{\alpha_1}{R} - 2\Lambda + R + \frac{R^2}{\beta^2} + \frac{R^3}{\beta^3} + \ldots.
\label{fR.series}
\end{equation}
Then, choosing various coefficients (sets of coefficients) one can find a number of phenomenologically interesting cases.

Among $f(R)$ theories there are many ones restricted by observable data and tests. Of course, researchers are interested in so called cosmological viable $f(R)$ models. In definition of \cite{1106.5582}, for example, they have to satisfy 1) the positivity of the effective gravitational coupling, $f'(R)>0$, that follows from the positivity of an effective gravitational constant $G_{eff} = G/f'(R) >0$; 2) the stability of cosmological perturbations, it is studied, for example, in \cite{Dolgov_2003} and references therein;  3) the existence of a matter-dominated stage;
4) the stability of the de-Sitter expansion at the present epoch, see, for example, \cite{Amendola.1};
5) the Solar System constraints (see, for example, \cite{Chiba.1,Chiba.2,BG1} and references therein), only they become consistent with $f(R)$ theory under chameleon mechanism; and 6) the constraint from the violation of the equivalence principle that becomes consistent under chameleon mechanism as well, see, for example, \cite{Capozziello}.

Each of the aforementioned possible tests for restrictions of $f(R)$ is used for criticism of the various models of $f(R)$ theories, see, for example, \cite{Dolgov_2003,Amendola.2}.
Nevertheless, it was constructed theories which can be classified as cosmological viable $f(R)$ gravities, for example, Starobinsky's model \cite{Starobinsky}, exponential gravity  \cite{0712.4017,Linder,1012.2280}, models of $f(R)$ cosmic acceleration \cite{Hu}. Thus, one can be convinced that $f(R)$ gravity is studied more intensively than many other modified theories preserving good perspectives for a development. By this, analysis of $f(R)$ theories in various aspects is quite desirable, and, here, we follow this aspiration considering gravitational waves in $f(R)$ gravity.

A necessity of the existence of gravitational waves in GR has been asserted by Einstein's work \cite{Einstein} immediately after constructing GR. During the following century, a theoretical study of gravitational waves in GR has being carried out very intensively.  Nowadays, after detecting gravitational waves by LIGO and Virgo antennas \cite{Abbott1,Abbott2,Abbott3} a new branch of cosmic observations and studies, namely, gravitational wave astronomy, starts and actively is developing.
Then, it is natural that  the question of the study of gravitational waves in the modified theories is increased sharply. At the nowadays the above observations are compatible with GR predictions \cite{Abbott1,Abbott4,Abbott5}. This may
help us to rule out or limit corresponding predictions of alternative gravity theories within the limits of the detectors.
In another word, at the enough exact level of observations one can obtain real restrictions to such theories. Therefore, it is important to understand the predictions of the
alternative theories with respect to gravitational waves.

To fill the list of numerous modified theories where gravitational waves take a place it could be useful to turn to the review \cite{CFPS1}, and to  references therein as well.
Concerning gravitational waves in $f(R)$ gravity, various their aspects are studied, for example,
in \cite{1106.5582,BG1,1608.01764,1701.05998,2411.06706,1907.06919,2205.07048,2108.11231,2112.02584,2204.00876,2112.15059,2409.07702}.


Isaacson \cite{A1,A2} was the first who has considered weak gravitational waves in high frequency limit and has demonstrated
a possibility to construct for them an averaged energy-momentum tensor. The latter bends a spacetime, and, as a result, gravitational waves are propagated on a curved background induced by a back-reaction. Isaacson's method and its generalizations were developed and applied in many of modified theories, for example, in \cite{SY1,PM1,P1,LC1,DLLL1}. Thus, in \cite{SY1}, especial attention is paid to  Chern-Simons gravity; in \cite{PM1} and \cite{P1} authors discuss higher-derivative theories for which cosmological back-reaction and its implications for a representation of dark energy are studied; in \cite{LC1} and \cite{DLLL1} reduced Horndeski theories and vector-tensor theories, respectively, are considered.

Among all the aforementioned papers we note the paper \cite{BG1} especially. Authors study linearized $f(R)$ gravity describing gravitational waves and test the results applying them in Solar System.  The function $f(R)$ is assumed as analytic about $R = 0$, therefore it is expanded in the form $f(R) = R + \alpha R^2 + \beta R^3 + \ldots $. One of the important results in  \cite{BG1} is a construction of  the averaged energy-momentum tensor for gravitational waves following Isaacson's recommendations. From the beginning authors consider perturbations on the flat (Minkowski) background for that it is enough to restrict the function to $f(R) = R + \alpha R^2$. Under such conditions they demonstrate that perturbations decompose onto two different parts: the classical traceless gravitational waves moving with the light speed and a ``massive'' part with non-zero trace moving with the speed less than light speed. Finally, constructing energy-momentum tensor for gravitational waves propagating on a curved background induced by a back-reaction, authors derive classical Isaacson's term and an additional term, which appears due to the non-zero trace part of perturbations.

In the most of works, where in modified theories the Isaacson procedure was applied, the flat background was used. However, a derivation of gravitational waves in the presence of a cosmological constant can lead to significant modifications with respect to the usual results even in GR, see, for example, \cite{Bernabeu}.
Thus making the use of a cosmological background for derivation of perturbations is of most importance. Reflecting this discussion to modified theories we stress the importance of the de Sitter (dS) space in $f(R)$ gravity among them.
For the best of our knowledge, there are no works in $f(R)$ gravity where as a starting background for perturbations in the Isaacson scheme was chosen just the dS space.

Thus the goal of our study is to close this gap developing the results of \cite{BG1} on the dS background.
We check a possibility of a generalization of the results \cite{BG1} in constructing energy-momentum tensor for gravitational waves and realize these possibilities. From the start we assume an arbitrary, but permissible, form of $f(R)$ and arbitrary background spacetime which is a solution to the $f(R)$ theory. Of course, we assume that $f(R)$ theory is physically viable.  Then, step by step we restrict such a freedom. First, we realize a construction of linear equations of the $f(R)$ gravity on the dS background,
 note that effective cosmological constant is modelled by inner structure of $f(R)$, like in \cite{Starobinsky,Hu}, without external cosmological constant in Lagrangian.
Second, we construct the averaged on the dS background energy-momentum for gravitational waves. All of these is used to construct the averaged energy-momentum on the curved background  induced by a back-reaction.

In the initial version of our calculations, we have used the tensor form of the $f(R)$ only, however in this case calculations become very cumbersome. Besides, the variable corresponding the inevitable scalar massive mode has to be recognized by a not trivial way. On the other hand, calculations with the scalar-tensor form of $f(R)$ are more economical and the scalar massive mode appears automatically. Thus, due to these advantages, we prioritize to work under such
a representation. Nonetheless, whenever the calculations do not become
too tiring, we use both the scalar-tensor and pure tensor forms
interchangeably, as they are equivalent.     Thus, finally, the averaged energy-momentum tensor on the dS background
is presented both in the pure tensor form and in the scalar-tensor
form that once more stresses the fact that both of the approaches are
equivalent.

The paper is organized as follows.

In section \ref{sect.2}, we give the main formulae of the $f(R)$ theory representing it into the form of the scalar-tensor theory by the standard Legendre transformations.
We describe also metric and scalar perturbations defined on an arbitrary background, which is a solution to a vacuum $f(R)$ theory.
Besides, we perform series expansions of more important quantities up to second order of smallness.

In section \ref{sect.4}, the system of linear equations is analyzed in detail. To simplify it gauge freedoms induced by the diffeomorphism invariance of $f(R)$ theory  are used. Concretization of background spacetimes to the solutions named as the {\em Einstein spaces} \cite{PetrovAZbook} allows us to decouple the full system of the linear equations onto a system for the metric perturbations and the equation for the scalar perturbation separately. Finally, these equations are simplified when the background is restricted to the dS spaces.

In section \ref{AppendixC}, first, we give a cumbersome calculus  for the quadratic expression on an arbitrary background. Second, we calculate  the averaged on the dS background energy-momentum tensor for gravitational waves.

In section \ref{sect.5}, we outline the Isaacson procedure and develop it in the $f(R)$ theory. We consider two cases, 1) when a back-reaction is negligible with respect to the initial curvature of the dS space (preserving the results of sections \ref{sect.4} and \ref{AppendixC}); and 2) when a back-reaction prevails the curvature of the dS space. In both the cases, we present the averaged energy-momentum tensor for gravitational waves.

In section \ref{sect.6},
we discuss our results obtained on the initial dS background, compare them with the results constructed on the Minkowski background in \cite{BG1} and  give concluding remarks.

In Appendix A, we present necessary relations ensuring our results and discuss these relations.

\section{Preliminaries }
\label{sect.2}

\subsection{The scalar-tensor representation of $f(R)$ theory }
\label{s-t-repres}

The action for $f(R)$ theory in vacuum \cite{FT1} is
\begin{equation}
S=\frac{1}{16\pi G}\int d^4x \sqrt{-g}f(R),
\label{1.1}
\end{equation}
where $R$ is the curvature scalar constructed with the metric $g_{ik}$ in the usual way in a pseudo-Riemannian space-time; $f(R)$ is enough smooth scalar function; the units correspond to a choice $c=1$. The equations of motion (field equations) corresponding to (\ref{1.1}) are
\begin{equation}
f'R_{ik} -\frac{1}{2}fg_{ik} +g_{ik}\Box f' -\nabla_i\nabla_k f' =0.
\label{1.2}
\end{equation}
Here and below we use the Latin letters to denote  all space-time indices; prime means a differentiation with respect to argument; the Ricci tensor $R_{ik}$ and the covariant derivatives $\nabla_i$ are constructed with the use of $g_{ik}$; $\Box \equiv \nabla_i\nabla^i$.

The  equations (\ref{1.2}) are of the fourth order ones. By this reason, for our study it is more comfortable to rewrite the theory $f(R)$ by using the Legendre transformations \cite{CFPS1}. Let us introduce the action
\begin{equation}
S=\frac{1}{16\pi G}\int d^4x \sqrt{-g}[f(\chi) + f'(\chi)(R-\chi)],
\label{1.3}
\end{equation}
where $\chi$ is an auxiliary scalar field. Varying (\ref{1.3}) with respect to $\chi$ one obtains
$$
f''(\chi)(R-\chi)=0.
$$
For a non-degenerated case $f''\neq 0$ one has $\chi=R$, the case $f'' = 0$ corresponds to general relativity. Substituting $\chi=R$ into the action (\ref{1.3}) we reproduce the initial action (\ref{1.1}).
Now, let us define the scalar field $\phi$ by a relation
\begin{equation}
\phi\equiv f'(\chi).
\label{1.4}
\end{equation}
Implying that it is reversible, $\chi= \chi(\phi)$, we can rewrite (\ref{1.3}) as action depending on independent dynamic variables $g_{ik}$ and  $\phi$ as follows
\begin{equation}
S=\frac{1}{16\pi G}\int d^4x \sqrt{-g}[f(\chi(\phi)) + \phi(R-\chi(\phi))]=\frac{1}{16\pi G}\int d^4x \sqrt{-g}[ \phi R- 2U(\phi)]
\label{1.5}
\end{equation}
with the potential
\begin{equation}
U(\phi)=\frac{1}{2}[\phi\chi(\phi) - f(\chi(\phi))].
\label{1.6}
\end{equation}
Now, varying action (\ref{1.5})
with respect to $g^{ik}$ and  $\phi$, we find out the field equations in the form of tensor and scalar parts as
\begin{equation}
J_{ik}\equiv\phi G_{ik} + g_{ik}\Box\phi + U(\phi) g_{ik} -\nabla_i\nabla_k\phi = 0,
\label{a.1}
\end{equation}
\begin{equation}
\Phi\equiv R-2U_{\phi} = 0,
\label{1.8}
\end{equation}
respectively, where $G_{ik}\equiv R_{ik}-\frac{1}{2}g_{ik}R$ is the Einstein tensor and $U_{\phi}=dU/d\phi$. Combining the trace of (\ref{a.1}) with (\ref{1.8}) one obtains equation
\begin{equation}
\tilde\Phi\equiv3\Box\phi +4U(\phi)-2\phi U_{\phi}=0,
\label{a.2}
\end{equation}
which can be more convenient in calculation.
Using the relations (\ref{1.4}), (\ref{1.6}) and
\begin{equation}
R=\chi(\phi)
\label{a.2+}
\end{equation}
 one easily converts the equations (\ref{a.1}) and (\ref{a.2}) into the initial form of equation (\ref{1.2}) in the $f(R)$ theory.

\subsection{Expansions of main expressions in perturbations}
\label{sect.3}

In this section, we present a formal approximate scheme on basis of which we derive necessary expansions for all the items in the expressions presenting field equations  (\ref{a.1})-(\ref{a.2}) of $f(R)$ theory.
As a starting point in constructing perturbation scheme we rewrite (\ref{a.1})-(\ref{a.2}) in the background form:
\begin{equation}
J^{(0)}_{ik}\equiv\phi^{(0)} G^{(0)}_{ik} + g^{(0)}_{ik}\Box^{(0)}\phi^{(0)} +  g^{(0)}_{ik}U^{(0)} -\nabla^{(0)}_{i}\nabla^{(0)}_{k}\phi^{(0)} = 0,
\label{a.1_0}
\end{equation}
\begin{equation}
\Phi^{(0)}\equiv R^{(0)}-2U^{(0)}_{\phi} = 0,
\label{1.80}
\end{equation}
\begin{equation}
\tilde\Phi^{(0)} \equiv 3\Box^{(0)}\phi^{(0)} +4U^{(0)}-2\phi^{(0)} U^{(0)}_{\phi}=0.
\label{a.2_0}
\end{equation}
Here and below: $\nabla_m^{(0)}$ is covariant derivative constructed with $ g_{ik}^{(0)}$;  $\Box^{(0)} = \nabla^{n(0)}\nabla_{n}^{(0)}$; indices are raised and lowered by $ g^{mn\,(0)}$ and $ g_{mn}^{(0)}$.
We assume that the background equations  (\ref{a.1_0}), (\ref{1.80}) and (\ref{a.2_0}) hold that means that background quantities $g^{(0)}_{ik}$ and $\phi^{(0)}$ satisfy them.

Next, we decompose the dynamic metric $g_{ik}$ (with the lower indices) into the background $g_{ik}^{(0)}$ and perturbation $h_{ik}$ parts:
\begin{equation}
g_{ik}=g_{ik}^{(0)}+ h_{ik}.
\label{b.1}
\end{equation}
Following Isaacson  \cite{A1,A2}, in the decomposition (\ref{b.1}) we do not assume infinite series in perturbations. At the moment, we suppose $g_{ik}^{(0)} \sim O(1)$, whereas  $h_{ik}\ll g_{ik}^{(0)}$. Then the decomposition (\ref{b.1}) can be thought in the form
$
g_{ik}=g_{ik}^{(0)}+ \epsilon H_{ik}
$
as well, where $H_{ik} \sim O(1)$ with dimensionless parameter $\epsilon \ll 1$. Thus, $h_{ik} \sim \epsilon$ and we assume also
\begin{eqnarray}
h_{ik} \sim \partial_n h_{ik} \sim \partial_{m}\partial_{n} h_{ik}\sim \ldots \sim \epsilon.
\label{h_sim}
\end{eqnarray}
Due to our setup we will study expressions up to $\epsilon^2$ and neglect by the higher orders.
Keeping in mind the above, first, we derive the expansion of metric determinant $g = \det g_{ik}$ with respect to the background metric determinant $g^{(0)} = \det g^{(0)}_{ik}$:
\begin{equation}
g=g^{(0)}(1+ h+ \frac{1}{2}h^2 -\frac{1}{2}h^i_kh_i^k)+O(\epsilon^3),
\label{b.1b}
\end{equation}
where  $h=g^{ik(0)} h_{ik}$. Second, using the evident relation $ g^{il} g_{lk}= \delta^i_k$, we derive the expasion of the contravariant metric components:
\begin{equation}
g^{ik}= g^{ik\,(0)}- h^{ik} + h^{il}h_l^k + O(\epsilon^3).
\label{b.1.1}
\end{equation}

Now, let us define expansions for the scalar field:
\begin{equation}
\phi=\phi^{(0)}+\phi^{(1)}+\phi^{(2)}+ O(\epsilon^3),
\label{b.7ab}
\end{equation}
where $\phi^{(0)}\sim O(1)$, and we assume
\begin{eqnarray}
\phi^{(1)}\sim \partial_m \phi^{(1)}\sim \partial_{m}\partial_{n} \phi^{(1)}\sim \ldots\sim \epsilon,
\label{phi_sim}\\
\phi^{(2)}\sim \partial_m \phi^{(2)}\sim \partial_{m}\partial_{n} \phi^{(2)}\sim \ldots\sim \epsilon^2.
\label{phi2_sim}
\end{eqnarray}
Assumptions (\ref{h_sim}), (\ref{phi_sim}) and (\ref{phi2_sim}) tells us that a characteristic wavelength of the gravitational ripple $\lambda$ is chosen as $O(1)$. Whereas, derivatives of $g^{(0)}_{ik}$ and $\phi^{(0)}$ behave in correspondence with the background model defined by (\ref{a.1_0}) - (\ref{a.2_0}).

From the start we give linear and quadratic in perturbations expressions
for Christoffel symbols:
\begin{equation}
\Gamma^{i\,(1)}_{kl}=\frac{1}{2} g^{im\,(0)}(\nabla_l^{(0)}h_{mk}+\nabla_k^{(0)} h_{ml}-\nabla_m^{(0)} h_{kl}),
\label{b.6}
\end{equation}
\begin{equation}
\Gamma^{i\,(2)}_{kl}= -\frac{1}{2}h^{im}(\nabla_l^{(0)} h_{mk} + \nabla_k^{(0)} h_{ml} -\nabla_m^{(0)} h_{kl}),
\label{b.7}
\end{equation}
which are, as one can see, tensor quantities, unlike of the connection $\Gamma^{i\,(0)}_{kl}$. Next, we present linear and quadratic approximations for the Riemannian tensor through $\Gamma^{i\,(1)}_{kl}$ and $\Gamma^{i\,(2)}_{kl}$:
\begin{equation}
R^{i\,(1)}_{klm}=\nabla_l^{(0)}  \Gamma^{i\,(1)}_{km}- \nabla_m^{(0)}  \Gamma^{i\,(1)}_{kl},
\label{b.7.1}
\end{equation}
\begin{equation}
R^{i\,(2)}_{klm}=\nabla_l^{(0)}  \Gamma^{i\,(2)}_{km}- \nabla_m^{(0)}  \Gamma^{i\,(2)}_{kl}   + \Gamma^{i\,(1)}_{nl}\Gamma^{n\,(1)}_{km} -\Gamma^{i\,(1)}_{nm}\Gamma^{n\,(1)}_{kl}.
\label{b.7.2}
\end{equation}

First of all, to expand the Einstein tensor in (\ref{a.1}), one needs to expand Ricci tensor $R_{ik}=R_{ik}^{(0)}+R_{ik}^{(1)}+R_{ik}^{(2)}$ and Ricci scalar $R=R^{(0)}+R^{(1)}+R^{(2)}$ in $\epsilon$ orders. The background quantities $R_{ik}^{(0)}$ and $R^{(0)}$ are constructed with $ g_{ik}^{(0)}$ in the usual way. To derive $R_{ik}^{(1)}$ one has to contract indices in (\ref{b.7.1}) and use the structure (\ref{b.6}), then the well known expression \cite{LL,MTW} is obtained
\begin{equation}
R^{(1)}_{km}= \frac{1}{2}\Big[\nabla_{l}^{(0)} \nabla_{m}^{(0)} h^l_k +\nabla_{l}^{(0)} \nabla_{k}^{(0)} h^l_m -\Box^{(0)} h_{km} -\nabla_{m}^{(0)}\nabla_{k}^{(0)}h  \Big].
\label{b.2}
\end{equation}
Contracting expanded Ricci tensor with (\ref{b.1.1}) and preserving the first order in $\epsilon$ one obtains easily
\begin{equation}
R^{(1)}= \nabla_{l}^{(0)}\nabla_{m}^{(0)} h^{lm}-\Box^{(0)} h-h^{ik}R^{(0)}_{ik}.
\label{b.3}
\end{equation}
Analogously to the above calculations, although in a more complicated way, we get expressions of the second order
\begin{eqnarray}
R^{(2)}_{km}=&& -\frac{1}{2}\nabla_l^{(0)}\Big [ h^{lq} \big( \nabla_m^{(0)}h_{qk} + \nabla_k^{(0)}h_{qm} -\nabla_q^{(0)}h_{km} \big) \Big ]+\frac{1}{2} \nabla_m^{(0)}\Big [ h^{lq} \nabla_k^{(0)} h_{ql}  \Big ]\nonumber\\
&&+\frac{1}{4} \nabla_n^{(0)} h \big( \nabla_m^{(0)} h^n_{k} + \nabla_k^{(0)} h^n_{m} -\nabla^{n\,(0)} h_{km} \big)
\nonumber\\
&&
-\frac{1}{4}\big( \nabla_m^{(0)} h^l_{n} + \nabla_n^{(0)} h^l_{m} -\nabla^{l\,(0)}h_{nm} \big)\big( \nabla_l^{(0)} h^n_{k} + \nabla_k^{(0)} h^n_{l} -\nabla^{n\,(0)}h_{kl} \big)
\label{b.4}
\end{eqnarray}
and
\begin{eqnarray}
R^{(2)}=&&h^{km}\Box^{(0)} h_{km} +h^{mk}\nabla_{m}^{(0)}\nabla_{k}^{(0)} h -\frac{1}{4}\nabla_n^{(0)} h\nabla^{n\,(0)} h +\nabla_l^{(0)} h^{lq} \nabla_q^{(0)} h -\nabla_l^{(0)}h^l_q\nabla_m^{(0)}h^{mq} \nonumber\\
&& -h^{km}\nabla_{l}^{(0)}\nabla_{m}^{(0)}h^l_k  -h^{kl}\nabla_{l}^{(0)}\nabla_{m}^{(0)}h^m_k +\frac{3}{4}\nabla_l^{(0)} h^{mn}\nabla^{l\,(0)} h_{mn} -\frac{1}{2}\nabla_m^{(0)}h^l_n \nabla^{n\,(0)} h^m_l+h^{kn}h_n^mR^{(0)}_{km}.
\label{b.5}
\end{eqnarray}

Now, let us return to the scalar field. Due to (\ref{1.4}) $\phi$ in (\ref{b.7ab}) can be rewritten as
\begin{equation}
\phi=\phi^{(0)}+\phi^{(1)}+\phi^{(2)}=f'_0+f'^{(1)}+f'^{(2)}
\label{b.7a}
\end{equation}
with
\begin{eqnarray}
f'_0 &\equiv & f'(R^{(0)}),
\label{b.7a+}\\
f'^{(1)}&\equiv &f''_0R^{(1)},
\label{b.7a++}\\
f'^{(2)}&\equiv & f''_0R^{(2)} +\frac{1}{2}f'''_0(R^{(1)})^2.
\label{b.7a+++}
\end{eqnarray}
Due to (\ref{1.6}), (\ref{a.2+}) and (\ref{b.7a})-(\ref{b.7a+++}) decomposition for $U$ reads
\begin{eqnarray}
U(\phi)&=&\frac{1}{2}\Big[ (\phi^{(0)}+\phi^{(1)}+\phi^{(2)})(\chi^{(0)}+\chi^{(1)}+\chi^{(2)}) -(f^{(0)}+f^{(1)}+f^{(2)})  \Big]\nonumber\\
&=&\frac{1}{2}\Big[ f'_0 R^{(0)} -f_0\Big]+ \frac{1}{2}\Big[ f''_0R^{(1)}R^{(0)}   \Big]+ \frac{1}{2}\Big[  f''_0R^{(2)}R^{(0)} +\frac{1}{2}f'''_0R^{(0)}(R^{(1)})^2 +\frac{1}{2}f''_0(R^{(1)})^2  \Big].
\label{b.8a}
\end{eqnarray}
Decomposition for $U_{\phi}$ obviously reads
\begin{equation}
U_{\phi}=\frac{1}{2}R^{(0)}+\frac{1}{2}R^{(1)}+\frac{1}{2}R^{(2)}.
\label{b.8a+}
\end{equation}

For expansions of higher derivative terms we use the next expressions
\begin{eqnarray}
\nabla_{i}\nabla_{k} \phi &=& \left(\nabla_{i} \nabla_{k} \phi\right)^{(0)} + \left( \nabla_{i}\nabla_{k} \phi\right)^{(1)} + \left( \nabla_{i}\nabla_{k} \phi\right)^{(2)}\nonumber\\
&=&\Big(\nabla_{i}^{(0)}\nabla_{k}^{(0)} \phi^{(0)}\Big)+\Big(\nabla_{i}^{(0)}\nabla_{k}^{(0)} \phi^{(1)}
-\Gamma^{n\,(1)}_{ik}\nabla_n^{(0)}\phi^{(0)}
\Big) \nonumber\\
&&+\Big(\nabla_{i}^{(0)}\nabla_{k}^{(0)} \phi^{(2)} -\Gamma^{n\,(1)}_{ik}\nabla_n^{(0)}\phi^{(1)} -\Gamma^{n\,(2)}_{ik}\nabla_n^{(0)}\phi^{(0)}   \Big),
\label{a.3}
\end{eqnarray}
\begin{eqnarray}
g_{ik}\Box \phi
&=& \left( g_{ik}\Box \phi\right)^{(0)}+\left( g_{ik}\Box \phi\right)^{(1)}+\left( g_{ik}\Box \phi\right)^{(2)}\nonumber\\
&=& (g_{ik}^{(0)}+  h_{ik})(g^{pq\,(0)}- h^{pq} + h^{pl}h_l^q )\nabla_p\nabla_q \phi \nonumber\\
&=& \Big\{ g_{ik}^{(0)}g^{pq(0)}\left(\nabla_{p} \nabla_{q} \phi\right)^{(0)}  \Big\} + \Big\{  \Big(h_{ik}g^{pq\,(0)} -g_{ik}^{(0)}h^{pq}\Big)\left( \nabla_{p}\nabla_{q} \phi\right)^{(0)} + g_{ik}^{(0)}g^{pq(0)} \left( \nabla_{p}\nabla_{q} \phi\right)^{(1)}   \Big\} \nonumber\\
&& + \Big\{  \Big(g_{ik}^{(0)}h^{pl}h^q_l-h_{ik}h^{pl}\Big)\left( \nabla_{p}\nabla_{q} \phi\right)^{(0)} + \Big(h_{ik}g^{pq(0)}-g_{ik}^{(0)}h^{pq}\Big)\left(\nabla_{p} \nabla_{q} \phi\right)^{(1)} +
g_{ik}^{(0)}g^{pq(0)}\left( \nabla_{p}\nabla_{q} \phi\right)^{(2)}  \Big\}.
\label{a.4}
\end{eqnarray}

\section{Linear field equations on fixed backgrounds}
\label{sect.4}

In this section, we analyze linear in perturbations equations
\begin{equation}
J^{(1)}_{ik}\equiv \phi^{(0)} G^{(1)}_{ik}+\phi^{(1)} G^{(0)}_{ik}  + \big(g_{ik}\Box\phi\big)^{(1)}+U^{(0)}h_{ik}+U^{(1)}g_{ik}^{(0)}-\big(\nabla_i\nabla_k \phi\big)^{(1)}=0,
\label{a.1_1}
\end{equation}
\begin{equation}
\Phi^{(1)}\equiv R^{(1)}-2U^{(1)}_{\phi} = 0,
\label{1.8_1}
\end{equation}
\begin{equation}
\tilde\Phi^{(1)}\equiv 3\big(\Box\phi\big)^{(1)} +4 U^{(1)}- 2\phi^{(1)}U_{\phi}^{(0)}-2 \phi^{(0)}U_{\phi}^{(1)} =0,
\label{a.2_1}
\end{equation}
which are a linear approximation of (\ref{a.1}), (\ref{1.8}) and (\ref{a.2}).

 \subsection{Linear equations on an arbitrary background}

 First of all, it is well known that diffeomorphism invariance of an arbitrary covariant field theory induces a gauge invariance of the linear equations, see, for example,  \cite{PKLT}. Considering the equations (\ref{a.1_1}) and (\ref{1.8_1}) one concretizes a gauge transformations for linear perturbations as
\begin{eqnarray}
h_{ik} &=& h^{\dag}_{ik} + \pounds_\xi g^{(0)}_{ik} =  h^{\dag}_{ik} + \nabla^{(0)}_i \xi_k +\nabla^{(0)}_k \xi_i,
\label{gauge.h}\\
\phi^{(1)} &=& \phi^{{\dag}(1)} + \pounds_\xi \phi^{(0)} =  \phi^{{\dag}(1)} + \xi^i\nabla^{(0)}_i \phi^{(0)} ,
\label{gauge.phi}
\end{eqnarray}
where the displacement vector $\xi^i$ satisfies $\xi^i \sim \partial_m \xi^i \sim \partial_{mn} \xi^i \sim \ldots \sim \epsilon$; indices are lowered by the background metric $g^{(0)}_{ik}$. Then, substitution of (\ref{gauge.h}) and (\ref{gauge.phi}) into (\ref{a.1_1}) and (\ref{1.8_1}) gives
\begin{eqnarray}
J^{(1)}_{ik} &=& J^{{\dag}(1)}_{ik} + \pounds_\xi J^{(0)}_{ik},
\label{gauge.J}\\
\Phi^{(1)} &=& \Phi^{{\dag}(1)} + \pounds_\xi \Phi^{(0)} .
\label{gauge.Phi}
\end{eqnarray}
Then, because we set that the background equations (\ref{a.1_0}) and (\ref{1.80}) hold we can assert that the linear equations  (\ref{a.1_1}) and (\ref{1.8_1}) are invariant with respect to transformations (\ref{gauge.h}) and (\ref{gauge.phi}). This means that if $h_{ik}$ and $\phi^{(1)}$ satisfy equations (\ref{a.1_1}) and (\ref{1.8_1}) then $h_{ik}^{\dag}$ and $\phi^{\dag(1)}$ satisfy equations (\ref{a.1_1}) and (\ref{1.8_1}) as well. Thus, the set of variables $h_{ik},\phi^{(1)}$ and the set $h_{ik}^{\dag},\phi^{\dag(1)}$ are equivalent in the sense of the gauge transformations (\ref{gauge.h}) and (\ref{gauge.phi}).  Of course, all the linear combinations of the linear
equations (\ref{a.1_1}) and (\ref{1.8_1}) are gauge invariant with taking into account the background equations as well, for example, (\ref{a.2_1}).

Now we are in a position to study the system of tensor equation (\ref{a.1_1}) and the  scalar equation (\ref{a.2_1}), which have to be considered simultaneously. First, the equations (\ref{a.3}) and (\ref{a.4}) give
\begin{eqnarray}
\left( \nabla_{i}\nabla_{k} \phi\right)^{(1)}  =\nabla_{i}^{(0)}\nabla_{k}^{(0)} \phi^{(1)}
-\Gamma^{n\,(1)}_{ik}\nabla_n^{(0)}\phi^{(0)}
,
\label{a.3+}
\end{eqnarray}
\begin{eqnarray}
\left( g_{ik}\Box \phi\right)^{(1)}= h_{ik}\Box^{(0)}\phi^{(0)} -g_{ik}^{(0)}h^{pq}\nabla_p^{(0)}\nabla_q^{(0)}\phi^{(0)} +g_{ik}^{(0)}\Box^{(0)}\phi^{(1)} - g_{ik}^{(0)}g^{pq\,(0)}\Gamma^{n\,(1)}_{pq}\nabla_n^{(0)}\phi^{(0)} ,
\label{a.5}
\end{eqnarray}
respectively. Second, using (\ref{b.8a}) and (\ref{b.8a+}), one can easily check that the  equation (\ref{a.2_1}) acquires the form:
\begin{eqnarray}
\Box^{(0)}\phi^{(1)}   - g^{pq\,(0)}\Gamma^{n\,(1)}_{pq}\nabla_n^{(0)}f'_0 - h^{pq}\nabla_p^{(0)}\nabla_q^{(0)} f'_0+ \frac{1}{3}\Big(R^{(0)}-\frac{f'_0}{f''_0}\Big)\phi^{(1)}=0.
\label{a.2_1+}
\end{eqnarray}
Third, (\ref{a.2_0}), with using (\ref{b.8a}) and (\ref{b.8a+}) as well, gives
\begin{eqnarray}
\Box^{(0)}\phi^{(0)} = \frac{1}{3}\Big(2f_0-{f'_0}R^{(0)}\Big).
\label{a.20+}
\end{eqnarray}
A combination of (\ref{a.3+}), (\ref{a.5}), (\ref{a.2_1+}) and (\ref{a.20+}) and substitution into (\ref{a.1_1}) gives
\begin{eqnarray}
&&\phi^{(0)} R^{(1)}_{ik} -\frac{1}{2}\phi^{(0)} R^{(1)}g_{ik}^{(0)} -\frac{1}{2}\phi^{(0)} R^{(0)}h_{ik} +\phi^{(1)} R^{(0)}_{ik} -\frac{1}{2}\phi^{(1)} R^{(0)}g_{ik}^{(0)} +\frac{2}{3}f_0h_{ik}-\frac{1}{3}f'_0R^{(0)}h_{ik} +\frac{1}{3}g_{ik}^{(0)}f'_0R^{(1)}
\nonumber\\
\nonumber\\
&&
-\frac{1}{3}g_{ik}^{(0)}f''_0R^{{0}}R^{(1)} +\frac{1}{2}f'_0R^{(0)} h_{ik}-\frac{1}{2}f_0 h_{ik} +\frac{1}{2}f''_0R^{(0)}R^{(1)} g_{ik}^{(0)} -\nabla_i^{(0)}\nabla_k^{(0)} \phi^{(1)}
+\Gamma^{n\,(1)}_{ik}\nabla_n^{(0)}\phi^{(0)}  =0.
\label{a.10}
\end{eqnarray}
Next, keeping in mind  (\ref{b.7a}) and  (\ref{b.7a++}), and replacing $R^{(1)}=\phi^{(1)}/f''_0$, we split all terms in  (\ref{a.10}) onto two groups:
\begin{eqnarray}
&&\Bigg[f'_0 R^{(1)}_{ik}  +\frac{1}{6}f_0h_{ik}-\frac{1}{3}f'_0R^{(0)}h_{ik} +\Gamma^{n\,(1)}_{ik}\nabla_n^{(0)}f'_0\Bigg]
\nonumber\\
\nonumber\\
&& + \Bigg[ \phi^{(1)} R^{(0)}_{ik} -\nabla_i^{(0)}\nabla_k^{(0)} \phi^{(1)} -\frac{1}{6}\frac{f'_0}{f''_0} \phi^{(1)}g_{ik}^{(0)}  -\frac{1}{3}\phi^{(1)} R^{(0)}g_{ik}^{(0)} \Bigg]
=0.
\label{a.11}
\end{eqnarray}
Thus, the tensor equation (\ref{a.1_1}) transforms to the form (\ref{a.11}) and scalar equation (\ref{a.2_1}) transforms to the form (\ref{a.2_1+}).

Now let us redefine the tensor dynamic variable
\begin{eqnarray}
\bar{h}_{ik}= h_{ik} -\frac{1}{2}h g_{ik}^{(0)}   -b \phi^{(1)} g_{ik}^{(0)},
\label{a.13}
\end{eqnarray}
 where we assume $b$ as a function of order $O(1)$. Then because $\bar{h} = -h -4b \phi^{(1)}$ (\ref{a.13}) is easily converted:
\begin{eqnarray}
h_{ik}=\bar{h}_{ik}     -\frac{1}{2}\bar{h}g_{ik}^{(0)}   -b \phi^{(1)} g_{ik}^{(0)}.
\label{a.14}
\end{eqnarray}
For new variables the gauge transformations (\ref{gauge.h}) and (\ref{gauge.phi}) are rewritten as
\begin{eqnarray}
\bar h_{ik} &=& \bar h^{\dag}_{ik} + \nabla^{(0)}_i \xi_k +\nabla^{(0)}_k \xi_i - g_{ik}^{(0)}\nabla^{(0)}_n \xi^n -bg_{ik}^{(0)}\xi^n\nabla^{(0)}_n \phi^{(0)},
\label{gauge.hbar}\\
\phi^{(1)} &=&  \phi^{{\dag}(1)} + \xi^i\nabla^{(0)}_i \phi^{(0)} .
\label{gauge.phibar}
\end{eqnarray}
Thus, after exchanging $h_{ik}$ by $\bar h_{ik}$ in equations (\ref{a.11}) and (\ref{a.2_1+}) and substituting  (\ref{gauge.hbar}) and (\ref{gauge.phibar}) into them one finds that (\ref{a.11}) and (\ref{a.2_1+}) are left invariant with taking into account background equations. This means that we can use a freedom in definitions of 4 components of $\xi^k$ and restrict $\bar h_{ik}$ and $\phi^{(1)}$ by 4 constraints.
For our goals we choose them in the form of
 the Lorentz conditions:
\begin{eqnarray}
\nabla_k^{(0)}  \bar{h}^k_i=0,
\label{Lorentz}
\end{eqnarray}
restricting $\bar h_{ik}$ only. Of course, the Lorentz conditions can be implied in an arbitrary metric theory for metric perturbations on an arbitrary given spacetime background.

Let us start transformations from the tensor equation (\ref{a.11}) keeping in mind (\ref{Lorentz}). From the beginning we substitute (\ref{a.14}) into $R^{(1)}_{ik}$ in (\ref{b.2}). Then, first, we use the commuting standard relation \cite{LL}:
\begin{eqnarray}
 \left(\nabla_{l}^{(0)} \nabla_{m}^{(0)}-  \nabla_{m}^{(0)}\nabla_{l}^{(0)}  \right)\bar{h}^i_k= \bar{h}^i_n R^{n(0)}_{~kml} -\bar{h}^n_kR^{i(0)}_{~nml},
\label{commute}
\end{eqnarray}
second, we use the equalities $(\nabla_{l}^{(0)} \nabla_{m}^{(0)}-  \nabla_{m}^{(0)}\nabla_{l}^{(0)})\phi^{(1)}=(\nabla_{l}^{(0)} \nabla_{m}^{(0)}-  \nabla_{m}^{(0)}\nabla_{l}^{(0)})\bar h =0$ as applied to scalar quantities. As a result one obtains
\begin{eqnarray}
R^{(1)}_{ik}=\bar h^n_p R^{p(0)}_{~(ik)n} + \bar h^p_{(i}R^{(0)}_{k)p} - \frac{1}{2}\Box^{(0)} \Big(\bar{h}_{ik} - \frac{1}{2}g_{ik}^{(0)}\bar h\Big) + \nabla_{k}^{(0)}\nabla_{i}^{(0)}(b\phi^{(1)})
+ \frac{1}{2}g_{ik}^{(0)}\Box^{(0)}(b\phi^{(1)}).
\label{a.15}
\end{eqnarray}
Then we substitute (\ref{a.15}) into (\ref{a.11}),  use (\ref{b.6}) and (\ref{a.14}) to replace other terms in (\ref{a.11})  and obtain
\begin{eqnarray}
&&f'_0 \Big[\bar h^n_p R^{p(0)}_{~(ik)n} + \bar h^p_{(i}R^{(0)}_{k)p} - \frac{1}{2}\Box^{(0)} \Big(\bar{h}_{ik} - \frac{1}{2}g_{ik}^{(0)}\bar h\Big) +\nabla_{k}^{(0)} \nabla_{i}^{(0)}(b\phi^{(1)})
+ \frac{1}{2}g_{ik}^{(0)}\Box^{(0)}(b\phi^{(1)})\Big]
\nonumber\\
&& +\frac{1}{6}f_0\bar{h}_{ik}-\frac{1}{12}f_0\bar{h}g^{(0)}_{ik}-\frac{1}{6}f_0b\phi^{(1)}g^{(0)}_{ik}-\frac{1}{3}f'_0R^{(0)}\bar{h}_{ik} +\frac{1}{6}f'_0R^{(0)}\bar{h}g^{(0)}_{ik}+\frac{1}{3}f'_0R^{(0)}b\phi^{(1)}g^{(0)}_{ik}
\nonumber\\
\nonumber\\
&&  +\frac{1}{2}\Big[ \nabla^{(0)}_k\bar{h}_i^n+\nabla^{(0)}_i\bar{h}^n_k-\nabla^{n\,(0)}\bar{h}_{ik}\Big]\nabla_n^{(0)}f'_0 -\frac{1}{4}\Big[ \delta^n_k\nabla^{(0)}_i\bar{h} +\delta^n_i\nabla^{(0)}_k\bar{h} -g^{(0)}_{ik}\nabla^{n\,(0)}\bar{h} \Big]\nabla_n^{(0)}f'_0
\nonumber\\
\nonumber\\
&& -\frac{1}{2}\Big[ \delta^n_k\nabla^{(0)}_i b\phi^{(1)} +\delta^n_i\nabla^{(0)}_k b\phi^{(1)} -g^{(0)}_{ik}\nabla^{n\,(0)} b\phi^{(1)}\Big]\nabla_n^{(0)}f'_0
\nonumber\\
\nonumber\\
&& + \Bigg[ \phi^{(1)} R^{(0)}_{ik} -\nabla_i^{(0)}\nabla_k^{(0)} \phi^{(1)} -\frac{1}{6}\frac{f'_0}{f''_0} \phi^{(1)}g_{ik}^{(0)}  -\frac{1}{3}\phi^{(1)} R^{(0)}g_{ik}^{(0)} \Bigg]
=0.
\label{a.16.2}
\end{eqnarray}
The scalar equation (\ref{a.2_1+}) after making the use of (\ref{b.6}) and  substituting (\ref{a.14}) becomes
\begin{eqnarray}
\Box^{(0)}\phi^{(1)}  = -\frac{1}{3}\Big(R^{(0)}-\frac{f'_0}{f''_0}\Big)\phi^{(1)}
+ \bar{h}^{pq}\nabla_p^{(0)}\nabla_q^{(0)} f'_0 -\frac{1}{2}\bar{h}\Box^{(0)}f'_0 -b\phi^{(1)}\Box^{(0)}f'_0  + \nabla^{n\,(0)}(b \phi^{(1)}) \nabla_n^{(0)}f'_0=0.
\label{a.16.1}
\end{eqnarray}

\subsection{Linear equations on the dS background}

One can see that expressions in equations (\ref{a.16.2}) and (\ref{a.16.1}) are very cumbersome, therefore it is desirable to simplify them. The first of assumptions is that we set $R^{(0)}= {\rm const}$. Then,
the quantities $f_0,~f'_0,~f''_0,...$ become constants by definitions $f_0 = f(R^{(0)})$, $f'_0 = f'(R^{(0)})$, etc.
Important consequence is that the background scalar field becomes constant as well, see (\ref{b.7a}) and (\ref{b.7a+}),
\begin{equation}
\phi^{(0)} = f'_0 = {\rm const}.
\label{(A)dS_phi}
\end{equation}
Then, one has $\nabla_n^{(0)}f'_0=0$ and $\Box^{(0)}f'_0=0$ that simplifies (\ref{a.16.2}) and (\ref{a.16.1}) significantly. Besides for a simplification it is fruitful to chose $b=1/f'_0$. To define a constant $R^{(0)}$ one combines  equations (\ref{1.80}) and (\ref{a.2_0}) with  zero order of (\ref{b.8a}). Finally, the equation (\ref{a.1_0}) gives
\begin{equation}
R^{(0)}_{ik} =  \frac{1}{2}\frac{f_0}{f'_0}g^{(0)}_{ik}, \qquad R^{(0)} = 2 \frac{f_0}{f'_0}.
\label{(A)dS+}
\end{equation}
The solutions of these equations is a wide class named as the Einstein spaces \cite{PetrovAZbook}.

As a result, the scalar equation (\ref{a.16.1}) becomes\footnote{It is the equation for ``massive'' scalar mode that was described in detail in \cite{1106.5582} on the dS background and in \cite{BG1} on a flat background, therefore we do not discuss it here paying a more attention to study the averaged energy-momentum of gravitational waves. Of course, we are restricted by our primary conditions (\ref{phi_sim}) and (\ref{phi2_sim}).}
\begin{eqnarray}
\Box^{(0)}\phi^{(1)}  + \frac{2}{3}\Big(\frac{f_0}{f'_0}-\frac{f'_0}{2f''_0}\Big)\phi^{(1)} =0.
\label{a.16.1+}
\end{eqnarray}
Then, keeping in mind (\ref{(A)dS_phi})-(\ref{a.16.1+}) one finds that (\ref{a.16.2}) is simplified significantly as well,
\begin{equation}
\Box^{(0)} \Big(\bar{h}_{ik} - \frac{1}{2}g_{ik}^{(0)}\bar h\Big) - \frac{1}{2}\frac{f_0}{f'_0}g_{ik}^{(0)}\bar h - 2R^{(0)}_{mikn}\bar h^{mn}=0.
\label{a.16.4+++}
\end{equation}
Here, we stress that for backgrounds chosen as Einstein spaces \cite{PetrovAZbook}, which can be quite complicated, the system for linear perturbations are decoupled into separate equations for scalar perturbations (\ref{a.16.1+}) and for tensor perturbations (\ref{a.16.4+++}). Such a result was not evident from the start.

The next step is to choose $R^{(0)}_{imkn}$ in (\ref{a.16.4+++}). It is more interesting to consider dS backgrounds, and we set
\begin{equation}
R^{(0)}_{imkn} = \frac{1}{6}\frac{f_0}{f'_0}\Big(  g^{(0)}_{ik}g^{(0)}_{mn}- g^{(0)}_{mk}g^{(0)}_{in}\Big).
\label{(A)dS++}
\end{equation}
Then,  (\ref{a.16.4+++}) transforms to the equation
\begin{eqnarray}
\Box^{(0)} \Big(\bar{h}_{ik} - \frac{1}{2}g_{ik}^{(0)}\bar h\Big) -\frac{1}{3}\frac{f_0}{f'_0}\Big(\bar{h}_{ik} + \frac{1}{2}g_{ik}^{(0)}\bar h\Big)=0
\label{a.16.4}
\end{eqnarray}
the trace of that is
\begin{equation}
\Box^{(0)}\bar{h}+\frac{f_0}{f'_0}\bar{h}=0.
\label{a.16.4+}
\end{equation}
Then, the tensor equation acquires the form:
\begin{eqnarray}
\Box^{(0)}\bar{h}_{ik} -\frac{1}{3}\frac{f_0}{f'_0}\Big(\bar{h}_{ik} - g_{ik}^{(0)}\bar{h}\Big)=0.
\label{a.16.4++}
\end{eqnarray}

Let us return to the gauge transformations and gauge invariance. The transformations (\ref{gauge.hbar}) and (\ref{gauge.phibar}) on the Einstein space (and on the dS concretely) backgrounds with  (\ref{(A)dS_phi}) become
\begin{eqnarray}
\bar h_{ik} &=& \bar h^{\dag}_{ik} + \nabla^{(0)}_i \xi_k +\nabla^{(0)}_k \xi_i - g_{ik}^{(0)}\nabla^{(0)}_n \xi^n,
\label{gauge.hbar+}\\
\phi^{(1)} &=&  \phi^{{\dag}(1)}.
\label{gauge.phibar+}
\end{eqnarray}
To clarify out the presence of residual degrees of freedom one has to check the transformations (\ref{gauge.hbar+}) in the sense of preserving the Lorentz conditions (\ref{Lorentz}) and of a gauge invariance of the equations (\ref{a.16.4++}). Thus, substitution of (\ref{gauge.hbar+}) into (\ref{Lorentz}) and taking into account a commuting relation analogous to (\ref{commute}) one obtains the condition
\begin{eqnarray}
\Box^{(0)}\xi^{i}  + \frac{1}{2}\frac{f_0}{f'_0}\xi^i =0
\label{Box.xi}
\end{eqnarray}
that preserves (\ref{Lorentz}). Thus, in transformations (\ref{gauge.hbar+}), one has a possibility to use $\xi^i$ satisfying (\ref{Box.xi}) only. Now, let us substitute (\ref{gauge.hbar+}) into the tensor equation (\ref{a.16.4++}). Again, after numerous applications of commuting relations analogous to (\ref{commute}) one obtains the condition
\begin{eqnarray}
\nabla_k^{(0)}\Big(\Box^{(0)}\xi_{i}\Big)  + \nabla_i^{(0)}\Big(\Box^{(0)}\xi_{k}\Big) -g_{ik}^{(0)} \nabla^{(0)m}\Big(\Box^{(0)}\xi_{m}\Big) + \frac{1}{2}\frac{f_0}{f'_0}\Big(\nabla_k^{(0)}\xi_{i} + \nabla_i^{(0)}\xi_{k} -g_{ik}^{(0)} \nabla^{(0)m} \xi_{m} \Big) =0.
\label{Box.xi.3}
\end{eqnarray}
 One can see that a condition (\ref{Box.xi}) guaranties a fulfilment of (\ref{Box.xi.3}). By this one can assert that residual degrees of freedom exist and are restricted by (\ref{Box.xi})
 creating the gauge invariance of tensor equation (\ref{a.16.4++}).

One has to remark that the tensor equation (\ref{a.16.4++}) exactly repeats the equation for linear tensor perturbations under the Lorentz condition (\ref{Lorentz}) in general relativity with the Einstein cosmological constant on dS backgrounds, but without matter sources; see, for example, \cite{Higuchi_1991,Vega_1999,Ashtekar_2015,Date_2016} and references therein. Thus, we can apply methods of these works to simplify the equation  (\ref{a.16.4++}). It turns out that for the solutions $\bar h_{ik}$  satisfying (\ref{a.16.4++}) with (\ref{Lorentz}) there is a possibility to redefine them by the way when $\bar h =0$. We show this following \cite{Higuchi_1991}.

Let the components $\bar h^*_{ik}$ satisfy (\ref{a.16.4++}) and (\ref{a.16.4+}) with  (\ref{Lorentz}). Now, let us introduce the new variables by the way
\begin{eqnarray}
\bar{h}_{ik} = \bar{h}^*_{ik}  -\frac{1}{2}g_{ik}^{(0)}\bar{h}^*     -  \frac{f'_0}{f_0}\nabla_{i}^{(0)}\nabla_{k}^{(0)} \bar{h}^*  .
\label{h_star}
\end{eqnarray}
From here, first, one can easily see that
\begin{eqnarray}
\bar{h}=0 .
\label{h_trace}
\end{eqnarray}
Second, one can check that $\bar h_{ik}$ satisfies the Lorentz conditions (\ref{Lorentz}) as well. Third, owing to (\ref{h_trace}) one can set that the tensor equation for $\bar h_{ik}$ in (\ref{h_star}) is
\begin{eqnarray}
\Box^{(0)}\bar{h}_{ik} -\frac{1}{3}\frac{f_0}{f'_0}\bar{h}_{ik} =0.
\label{h_tracless}
\end{eqnarray}
Substituting here $\bar h_{ik}$ from (\ref{h_star}) one obtains  (\ref{a.16.4++}) for  $\bar h^*_{ik}$. One has to remark that it is possible a further redefinition of $\bar{h}_{ik} \rightarrow {\cal H}_{ik}$ to absorb the effect of expansion of the universe that gives $\Box^{(0)}{\cal H}_{ik} = 0$, for detail see \cite{1106.5582}.

At last, to finalize a description of gauge transformations of the linear in perturbations equations we remark that the condition (\ref{Lorentz}) is added by a traceless condition  (\ref{h_trace}). Then, the restrictions for the residual freedoms (\ref{Box.xi}) are to be added by
\begin{eqnarray}
\nabla^{(0)}_i \xi^i =0.
\label{Box.xi+}
\end{eqnarray}
The residual freedoms restricted by (\ref{Box.xi}) and (\ref{Box.xi+}) together can be used fruitfully. For example, in \cite{Higuchi_1991} it was shown in detail that one can choose $\bar h_{i0}=0$ that, in summary, defines a so called TT-gauge \cite{LL,MTW}.

The choice of the dS background in (\ref{(A)dS++}) allows us to introduce the characteristic length $\ell$ of the initial background. It is the dS radius, or radius of the cosmological horizon, by (\ref{(A)dS++}) it is defined as $\ell^2 = 6f'_0/f_0$. Thus, (\ref{(A)dS+}) and (\ref{(A)dS++}) can be rewritten as
 \begin{eqnarray}
 R^{(0)}_{imkn} = \frac{1}{\ell^2}\Big(  g^{(0)}_{ik}g^{(0)}_{mn}- g^{(0)}_{mk}g^{(0)}_{in}\Big),\qquad R^{(0)}_{ik} =  \frac{3}{\ell^2}g^{(0)}_{ik}, \qquad R^{(0)} = \frac{12}{\ell^2}.
\label{R0_ell}
\end{eqnarray}
This specific form of the Riemannian tensor and its contractions just are related with a space behaviour that is required for the cosmologically viable $f(R)$
theories, see discussion in Introduction.
The dS metric $g^{(0)}_{ik}$ and scalar field $\phi^{(0)}$ (satisfying (\ref{(A)dS_phi})) and their derivatives has a behaviour:
\begin{eqnarray}
g_{ik}^{(0)}& =& O(1);\qquad\partial_n g_{ik}^{(0)} = O\left( 1/\ell\right);\qquad \partial_{m}\partial_{n}  g_{ik}^{(0)}= O\left( 1/\ell^2\right); \ldots ,
\label{g0_sim+}\\
\phi^{(0)} &=& O(1);\qquad\partial_n \phi^{(0)} = \partial_{m}\partial_{n} \phi^{(0)} = \ldots = 0.
\label{phi0_sim+}
\end{eqnarray}
At last, the behaviour (\ref{g0_sim+}) allows us to derive for the background Christoffel symbols:
 \begin{eqnarray}
 \Gamma^{i\,(0)}_{kl} = O\left( 1/\ell\right);\qquad\partial_n \Gamma^{i\,(0)}_{kl} = O\left( 1/\ell^2\right);\qquad \partial_{m}\partial_{n}  \Gamma^{i\,(0)}_{kl}= O\left( 1/\ell^3\right); \ldots.
\label{Gamma0_ell}
\end{eqnarray}

Recall that a behaviour (\ref{h_sim}), (\ref{phi_sim}) and (\ref{phi2_sim}) corresponds to assumption that a characteristic wavelength of the gravitational ripple $\lambda$ is chosen as $\lambda\sim1$. Because $\ell$ presents a radius of the cosmological horizon one must to set that it is significantly more than the gravitational wavelength, thus $\ell\gg 1$. Then, considering the final linear equations (\ref{a.16.1+}) and (\ref{h_tracless}) and keeping in mind (\ref{g0_sim+}) - (\ref{Gamma0_ell}), one finds that their left hand sides are splitting onto three orders
 \begin{eqnarray}
 \rightarrow\qquad O\left(\epsilon\right)  + O\left( \epsilon/\ell\right) +O\left( \epsilon/\ell^2\right),
\label{splitting}
\end{eqnarray}
where the leading order is determined by $g^{mn\,{(0)}}\partial_{m}\partial_{n}\bar h_{ik} \sim g^{mn\,{(0)}}\partial_{m}\partial_{n} \phi^{(1)} \sim \epsilon $. One has to remark that in (\ref{a.16.1+}) we suppose $f'_0/f''_0 \sim 1/\ell^2$ that is discussed in Appendix A.

\section{Averaged energy-momentum for gravitational waves on the dS background }
\label{AppendixC}

We recall that a one of the main goals of the paper is to construct the energy-momentum tensor for gravitational waves in $f(R)$ theory with taking into account a back-reaction. However, we consider a back-reaction in the next section only. In this section, we construct the energy-momentum on a fixed dS background introduced above that is quite instructive for the further study in the paper.

\subsection{Expression  $J^{(2)}_{ik}$ on an arbitrary background }
\label{AppendixD}

By the generally accepted notions in derivation of gravitational waves, their energy-momentum is defined by a second order of expansions of the field equations.
In our model the quantity $J^{(2)}_{ik}$ presenting the second order in the expansion of (\ref{a.1}) just has to play a role of effective energy-momentum. Its general structure is
\begin{eqnarray}
J^{(2)}_{ik} = \phi^{(0)} G^{(2)}_{ik}+\phi^{(2)} G^{(0)}_{ik} +\phi^{(1)} G^{(1)}_{ik} +h_{ik}\big(\Box\phi\big)^{(1)} +g_{ik}^{(0)}\big(\Box\phi\big)^{(2)}+U^{(1)}h_{ik}+U^{(2)}g_{ik}^{(0)}-\big(\nabla_{i} \nabla_{k}\phi\big)^{(2)}.
\label{a.12}
\end{eqnarray}
 Also for our calculations it is necessary the second order of the equation (\ref{a.2}) to take into account it (\ref{a.12}):
\begin{eqnarray}
\tilde\Phi^{(2)} = 3\left(\Box\phi \right)^{(2)} - 2\phi^{(2)}U^{(0)} - 2\phi^{(0)}U^{(2)} - 2\phi^{(1)}U^{(1)} + 4U^{(2)} = 0.
\label{Phi.2}
\end{eqnarray}

In this subsection, we derive the structure of $J^{(2)}_{ik}$ on an arbitrary background  keeping in mind the results in section \ref{sect.3}.
For the sake of convenience we represent all the terms in (\ref{a.12}) separately. In final expression (\ref{a.21}) we use background quantities $ f'_0,\,f''_0,\,f'''_0$ and $g_{ik}^{(0)},~\phi^{(0)}$; terms of expansions of Ricci tensor and curvature scalar: $R^{(0)}_{ik},\,R^{(1)}_{ik},\,R^{(2)}_{ik}$ and $R^{(0)},\,R^{(1)},\,R^{(2)}$, respectively; and perturbations $h_{ik}$, $\phi^{(1)}$ and $\phi^{(2)}$.

Thus, recalling the definition of the Einstein tensor, and using (\ref{b.1}) and (\ref{b.7a}), one has for the first term in (\ref{a.12}):
\begin{eqnarray}
\phi^{(0)} G^{(2)}_{ik}= f'_0 R^{(2)}_{ik} -\frac{1}{2} f'_0 h_{ik}  R^{(1)}   -\frac{1}{2} f'_0 g^{(0)}_{ik}  R^{(2)}.
\label{I.1}
\end{eqnarray}
Referring the formula (\ref{b.7a}) with (\ref{b.7a+++}) and recalling the definition of the Einstein tensor again one has for the second term in (\ref{a.12}):
\begin{eqnarray}
\phi^{(2)} G^{(0)}_{ik} = f''_0 R^{(0)}_{ik} R^{(2)} - \frac{1}{2} f''_0  g^{(0)}_{ik} R^{(0)} R^{(2)}+  \frac{1}{2}f'''_0 R^{(0)}_{ik}  (R^{(1)})^2 -\frac{1}{4} f'''_0  g^{(0)}_{ik} R^{(0)}(R^{(1)})^2  .
\label{I.2}
\end{eqnarray}
Referring the formula (\ref{b.7a}) with (\ref{b.7a++})  one has for the third term in (\ref{a.12}):
\begin{eqnarray}
\phi^{(1)} G^{(1)}_{ik} = f''_0 R^{(1)}_{ik} R^{(1)} -\frac{1}{2} f''_0  h_{ik} R^{(0)} R^{(1)} -  \frac{1}{2} f''_0 g^{(0)}_{ik} (R^{(1)})^2.
\label{I.3}
\end{eqnarray}
Substituting related terms from (\ref{b.7a}) with (\ref{b.7a++}), (\ref{b.8a}) and (\ref{b.8a+}) into (\ref{a.2_1}) one obtains for the fourth term in (\ref{a.12}):
\begin{eqnarray}
h_{ik}\big(\Box\phi\big)^{(1)} = \frac{1}{3} f'_0 h_{ik} R^{(1)} -\frac{1}{3} f''_0h_{ik} R^{(0)}  R^{(1)}.
\label{I.4}
\end{eqnarray}
To derive the fifth term in (\ref{a.12}) one has to use the equation (\ref{Phi.2}).
Substituting there the related terms from (\ref{b.7a}) with (\ref{b.7a+++}), (\ref{b.8a}) and (\ref{b.8a+}) one obtains for the fifth term in (\ref{a.12}):
\begin{eqnarray}
g_{ik}^{(0)}\big(\Box\phi\big)^{(2)} = \frac{1}{3} f'_0  g_{ik}^{(0)} R^{(2)} -\frac{1}{3} f''_0  g_{ik}^{(0)} R^{(0)} R^{(2)} -  \frac{1}{6}f'''_0 g^{(0)}_{ik} R^{(0)} (R^{(1)})^2.
\label{I.5}
\end{eqnarray}
The sixth and seventh terms in (\ref{a.12}) are defined by the related orders in (\ref{b.8a}):
\begin{eqnarray}
U^{(1)}h_{ik}=\frac{1}{2} f''_0  h_{ik} R^{(0)} R^{(1)},
\label{I.6}
\end{eqnarray}
\begin{eqnarray}
U^{(2)}g_{ik}^{(0)}=\frac{1}{2} f''_0 g^{(0)}_{ik} R^{(0)}  R^{(2)} +\frac{1}{4}f'''_0  g^{(0)}_{ik} R^{(0)} (R^{(1)})^2 +\frac{1}{4}f''_0 g^{(0)}_{ik} (R^{(1)})^2.
\label{I.7}
\end{eqnarray}
At last, the eighth  term in (\ref{a.12}) is determined by the second order term in (\ref{a.3})
\begin{eqnarray}
-\big(\nabla_{i}\nabla_{k} \phi\big)^{(2)} = -\nabla_{i}^{(0)}\nabla_{k}^{(0)}\phi^{(2)}+ f''_0\Gamma^{n\,(1)}_{ik}\nabla_n^{(0)} R^{(1)} +\Gamma^{n\,(2)}_{ik}\nabla_n^{(0)} \phi^{(0)}.
\label{I.8}
\end{eqnarray}
Thus, summing (\ref{I.1}) - (\ref{I.8}) one obtains
\begin{eqnarray}
J^{(2)}_{ik} =&&
f'_0 R^{(2)}_{ik} +  f''_0 R^{(1)}_{ik} R^{(1)}
- \frac{1}{6} f'_0 h_{ik}  R^{(1)} -\frac{1}{3} f''_0 R^{(0)} h_{ik} R^{(1)}
-\frac{1}{6} f'_0 g^{(0)}_{ik}  R^{(2)}
\nonumber\\
&&+f''_0 R^{(0)}_{ik} R^{(2)}  -\frac{1}{3} f''_0 R^{(0)} g_{ik}^{(0)} R^{(2)}
+  \frac{1}{2} R^{(0)}_{ik}f'''_0  (R^{(1)})^2 -  \frac{1}{6} g^{(0)}_{ik}f'''_0 R^{(0)} (R^{(1)})^2
\nonumber\\
&&-\frac{1}{4} g^{(0)}_{ik}f''_0 (R^{(1)})^2-\nabla_{i}^{(0)}\nabla_{k}^{(0)}\phi^{(2)}+ f''_0\Gamma^{n\,(1)}_{ik}\nabla_n^{(0)} R^{(1)}  +\Gamma^{n\,(2)}_{ik}\nabla_n^{(0)} \phi^{(0)}.
\label{a.21}
\end{eqnarray}

Let us discuss a role that can be played by $J^{(2)}_{ik}$ derived in (\ref{a.21}). Expand the equation (\ref{a.1}) as
\begin{eqnarray}
J^{(0)}_{ik}+J^{(1)}_{ik}+J^{(2)}_{ik}+J^{(3)}_{ik}+\ldots = 0.
\label{J.1.2.3}
\end{eqnarray}
Because now we consider a background as a fixed one the first term in (\ref{J.1.2.3}) disappears by (\ref{a.1_0}). The linear part in (\ref{J.1.2.3}) allows us to find out perturbations $h_{ik}$ and $\phi^{(1)}$ in the linear approximation by the equations (\ref{a.1_1}), together with (\ref{a.2_1}). The quadratic expression (\ref{a.21}) placed into the equation (\ref{J.1.2.3}) allows us to find corrections to $h_{ik}$ and $\phi^{(1)}$ of the next order by the equations
\begin{eqnarray}
J^{(1)}_{ik}(h_{ik}^{(2)},\phi^{(2)}) = - J^{(2)}_{ik}(h_{ik},\phi^{(1)}).
\label{J.1+2}
\end{eqnarray}
Of course, in analogous way the scalar type equation has to be taken into account.
The iteration procedure can be continued, thus, $J^{(3)}_{ik}$ can by applied to calculate the next corrections $h_{ik}^{(3)}$ and $\phi^{(3)}$, etc.

Unlike the above, in our consideration, it is important another role $J^{(2)}_{ik}$ when it is classified as an effective energy-momentum of gravitational waves in $f(R)$ theory. Then, such an energy-momentum has to influence onto a curvature of a spacetime. However, due to the Einstein equivalence principle gravitational energy is not localizable \cite{MTW}. Therefore, the energy-momentum has to be considered in a finite spacetime volume and the averaging procedure can be applied. We do it in the next subsection.

\subsection{The averaged energy-momentum on the dS background }
\label{AppendixB}

Here, we provide the averaging procedure for the energy-momentum (\ref{a.21}) on the dS background. In fact, under the averaging we have a possibility to take into account all the orders after the splitting on the dS background. However, we preserve only the leading order of the averaged energy-momentum cancelling step by step other orders. We use the rules of the Brill-Hartle averaging \cite{BH1} for regions with a scale $S$ significantly more than the characteristic length of the wavelength $\lambda \sim 1$ of the gravitational ripple $S \gg O(1)$, of course, $S \leq \ell$. It is important to recall that
 because the averaging is provided over all directions at each point in regions of the scale $S$, gradients average to zero \cite{BH1}:
\begin{eqnarray}
\langle \partial_m{\cal A}  \rangle = 0,\qquad \langle{\cal B}\partial_m{\cal A}\rangle = -\langle{\cal A}\partial_m{\cal B}\rangle.
\label{J.AAA}
\end{eqnarray}

Let us rewrite (\ref{a.21}) in a more convenient form: First, by (\ref{b.7a}) with (\ref{b.7a++}) we set $R^{(1)} = \phi^{(1)}/f''_0$ that gives a possibility to consider $R^{(1)}$ as independent quantity expressed through $\phi^{(1)}$. Second,  with the use of (\ref{(A)dS+})  we replace the background Ricci tensor by the background curvature scalar. Thus, (\ref{a.21}) is rewritten in the form:
\begin{eqnarray}
J^{(2)}_{ik} =&& - \frac{1}{6}f'_0 g_{ik}^{(0)}R^{(2)}\left[1+ \frac{1}{2} \frac{f''_0}{f'_0}R^{(0)} \right]
- \frac{1}{4}\frac{1}{f''_0} g_{ik}^{(0)}\left( \phi^{(1)}\right)^2\left[1+ \frac{1}{6} \frac{f'''_0}{f''_0}R^{(0)} \right]
- \frac{1}{3}\frac{f'_0}{f''_0} h_{ik} \phi^{(1)}\left[\frac{1}{2}+ \frac{f''_0}{f'_0}R^{(0)} \right]
\nonumber\\
&&
+  R^{(1)}_{ik} \phi^{(1)} + \Gamma^{n\,(1)}_{ik}\nabla_n^{(0)}\phi^{(1)}
+ f'_0 R^{(2)}_{ik} -\nabla_{i}^{(0)}\nabla_{k}^{(0)}\phi^{(2)} +\Gamma^{n\,(2)}_{ik}\nabla_n^{(0)} \phi^{(0)}.
\label{a.21+}
\end{eqnarray}
To analyze each of items in (\ref{a.21+}) one has to reset $h_{ik}$ by $\bar h_{ik}$ with making the use of (\ref{a.14}) that satisfies Lorentz conditions (\ref{Lorentz}) and traceless condition (\ref{h_trace}) with $b= 1/f'_0$:
\begin{eqnarray}
h_{ik}=\bar{h}_{ik}       - g_{ik}^{(0)}\phi^{(1)}/f'_0, \qquad h=      - 4\phi^{(1)}/f'_0.
\label{a.14+}
\end{eqnarray}
Owing to our assumptions of smallness in (\ref{h_sim}) and (\ref{phi_sim}) one has $f'_0 \sim O(1)$.

Keeping in mind (\ref{R0_ell}) and (\ref{prime}), one concludes that expressions in (\ref{a.21+}) in the first line in the square brackets have a behaviour $O(1)$. Therefore it is necessary to consider coefficients at these square brackets. Thus, let us consider the first term in (\ref{a.21+}). After substitution (\ref{a.14+}) into (\ref{b.5})  one has for the second order of Ricci scalar:
\begin{eqnarray}
R^{(2)}[\bar{h}_{ik}]=  -\bar{h}^{km}\nabla_l^{(0)}\nabla_m^{(0)}\bar{h}^l_k  +\frac{3}{4}\nabla_l^{(0)} \bar{h}^{mn}\nabla^{l\,(0)} \bar{h}_{mn} -\frac{1}{2}\nabla_m^{(0)}\bar{h}^l_n \nabla^{n\,(0)} \bar{h}^m_l+\bar{h}^{kn}\bar{h}_n^mR^{(0)}_{km}     \nonumber\\
+\frac{6}{f'^2_0} \phi^{(1)} \Box^{(0)}  \phi^{(1)}  -\frac{3}{2} \frac{1}{f'^2_0}  \nabla_n^{(0)}\phi^{(1)}\nabla^{n\,(0)}\phi^{(1)} -\frac{2}{f'_0} \bar{h}^{km}\nabla_m^{(0)}\nabla_k^{(0)}\phi^{(1)}
-\frac{2}{f'_0}\phi^{(1)}\bar{h}^{km} R^{(0)}_{km} +\frac{1}{f'^2_0} (\phi^{(1)})^2R^{(0)} .
\label{a.22}
\end{eqnarray}
On the example of this expression we outline the procedure of the averaging  in detail. For the next items in (\ref{a.21+}) we will derive the result of the averaging only.

For the first term in (\ref{a.22}), one uses the commuting relation (\ref{b.3}), then one applies the Lorentz condition (\ref{Lorentz}) and behaviour (\ref{R0_ell}). Finally one obtains
$$
\langle\bar{h}^{km}\nabla_l^{(0)}\nabla_m^{(0)}\bar{h}^l_k\rangle =O\left( \epsilon^2/\ell^2\right),
$$
where we recall $\ell \gg 1$.

For the second term in (\ref{a.22}),
\begin{eqnarray*}
\nabla_l^{(0)} \bar{h}^{mn}\nabla^{l\,(0)} \bar{h}_{mn} = \nabla_l^{(0)}\left(\bar{h}^{mn}\nabla^{l\,(0)} \bar{h}_{mn} \right) - \bar{h}^{mn}\Box^{(0)} \bar{h}_{mn}\\
= \partial _l\left(\bar{h}^{mn}\nabla^{l\,(0)} \bar{h}_{mn} \right) + \Gamma^{k\,(0)}_{kl}\bar{h}^{mn}\nabla^{l\,(0)}\bar{h}_{mn} - \frac{1}{3}\frac{f_0}{f'_0}\bar{h}^{mn}\bar{h}_{mn}.
\end{eqnarray*}
Here, the first term disappears under the averaging due to (\ref{J.AAA}), the second term has a behaviour $O\left(  \epsilon^2/\ell\right)$ by (\ref{Gamma0_ell}),  the third term is obtained due to (\ref{h_tracless}) and has the behaviour $O\left(  \epsilon^2/\ell^2\right)$, see (\ref{splitting}). Thus, finally
$$
\langle \nabla_l^{(0)} \bar{h}^{mn}\nabla^{l\,(0)} \bar{h}_{mn} \rangle = O\left(  \epsilon^2/\ell\right).
$$

For the third term in (\ref{a.22}) one replaces covariant derivatives again. Then, step by step one uses (\ref{J.AAA}), (\ref{R0_ell}), the commuting relation (\ref{b.3}) and the Lorentz conditions (\ref{Lorentz}). For the fourth term in (\ref{a.22}) one uses (\ref{R0_ell}). Finally one obtains
 \begin{eqnarray*}
 \langle \nabla_m^{(0)}\bar{h}^l_n \nabla^{n\,(0)} \bar{h}^m_l\rangle = O\left(\epsilon^2/\ell\right),\\
 \langle \bar{h}^{kn}\bar{h}_n^mR^{(0)}_{km} \rangle =  O\left(\epsilon^2/\ell^2\right).
 \end{eqnarray*}

For the fifth and sixth terms in (\ref{a.22}), first, one replaces covariant derivatives in the sixth item and, then, uses  the equation (\ref{a.16.1+}). Next, step by step one has to use  (\ref{J.AAA}), the relations of smallness (\ref{phi_sim}) and (\ref{R0_ell}). As a result, one obtains finally
 \begin{eqnarray*}
 \langle \phi^{(1)} \Box^{(0)}  \phi^{(1)}\rangle = O\left(\epsilon^2/\ell^2\right),\\
 \langle \nabla_n^{(0)}\phi^{(1)}\nabla^{n\,(0)}\phi^{(1)}   \rangle =  O\left(\epsilon^2/\ell\right).
 \end{eqnarray*}

The analogous arguments and the Lorentz conditions (\ref{Lorentz}) give the behaviour for the seventh  term in (\ref{a.22})
$$
\langle \bar{h}^{km}\nabla_m^{(0)}\nabla_k^{(0)}\phi^{(1)} \rangle=  O\left(\epsilon^2/\ell\right).
$$

At last, it is evidently that for last two terms in (\ref{a.22}) one has
$$
\langle -2\phi^{(1)}\bar{h}^{km} R^{(0)}_{km} + (\phi^{(1)})^2R^{(0)} \rangle = O\left(\epsilon^2/\ell^2\right).
$$
Finally, one has for (\ref{a.22}):
\begin{eqnarray}
\Big\langle - \frac{1}{6}f'_0 g_{ik}^{(0)} R^{(2)}[\bar{h}_{ik}]\Big\rangle =   O\left(\frac{\epsilon^2}{\ell}\right) .
\label{a.24}
\end{eqnarray}

Let us return to (\ref{a.21+}). In order to give a behaviour of the second and third terms in the first line of (\ref{a.21+}) we use the behaviour of ${f'_0}/{f''_0}$ and ${f''_0}/{f'''_0}$ given in (\ref{prime}) in Appendix A. The last restricts a class of functions $f(R)$ although it leaves enough arbitrary, see discussion in Appendix A. Thus,
\begin{eqnarray}
\Big\langle \frac{{f'_0}}{f''_0} \frac{{1}}{f'_0} g_{ik}^{(0)}\left( \phi^{(1)}\right)^2 \Big\rangle  \sim \Big\langle \frac{f'_0}{f''_0} h_{ik} \phi^{(1)}\Big\rangle = O\left(\frac{\epsilon^2}{\ell^2}\right).
\label{a.24+}
\end{eqnarray}
The fourth term in the second line of (\ref{a.21+}) after averaging has a behaviour:
\begin{eqnarray}
\Big\langle - \nabla_{i}^{(0)}\nabla_{k}^{(0)}\phi^{(2)} \Big\rangle   = \Big\langle -\partial_{i}\partial_{k} \phi^{(2)} + \Gamma^{n\,{(0)}}_{ik}\phi^{(2)} \Big\rangle = O\left(\frac{\epsilon^2}{\ell}\right).
\label{phi2.average}
\end{eqnarray}
The last term in the second line of (\ref{a.21+}) disappers for the dS background due to (\ref{(A)dS_phi}).

Now let us substitute (\ref{a.14+}) into the first two terms in (\ref{a.21+}) in the second line and derive them:
\begin{eqnarray}
R^{(1)}_{ik} \phi^{(1)}[\bar{h}_{ik}]&=& \frac{1}{2}\phi^{(1)}\left[\nabla_{l}^{(0)}\nabla_{i}^{(0)}\bar h^l_k + \nabla_{l}^{(0)} \nabla_{k}^{(0)}\bar h^l_i - \Box^{(0)}\left( \bar h_{ik} - g^{(0)}_{ik}\phi^{(1)}/f'_0\right)+ 2\nabla_{i}^{(0)}\nabla_{k}^{(0)}\phi^{(1)}/f'_0 \right],
\label{a.25}\\
\Gamma^{n\,(1)}_{ik}\nabla_n^{(0)} \phi^{(1)}[\bar{h}_{ik}]&=& \frac{1}{2}\left[\nabla_k^{(0)}\bar{h}_i^n + \nabla_i^{(0)}\bar{h}_k^n - \nabla^{n\,(0)}\bar{h}_{ik} \right]\nabla_n^{(0)} \phi^{(1)}\nonumber\\
 &&-\nabla_i^{(0)} \phi^{(1)} \nabla_k^{(0)} \phi^{(1)} /f'_0 +\frac{1}{2} g^{(0)}_{ik}\nabla^{n\,(0)} \phi^{(1)} \nabla_n^{(0)} \phi^{(1)}/f'_0 .
\label{a.27}
\end{eqnarray}

Applying to (\ref{a.25}) and  (\ref{a.27}) all the arguments used above for the averaging one obtains
\begin{eqnarray}
\Big\langle R^{(1)}_{ik} \phi^{(1)}[\bar{h}_{ik}]\Big\rangle &= & -\frac{1}{f'_0} \nabla_{i}^{(0)}\phi^{(1)} \nabla_{k}^{(0)}\phi^{(1)} + O\left(\frac{\epsilon^2}{\ell}\right),
\label{a.25+}\\
\Big\langle\Gamma^{n\,(1)}_{ik}\nabla_n^{(0)} \phi^{(1)}[\bar{h}_{ik}]\Big\rangle &=&
-\frac{1}{f'_0} \nabla_{i}^{(0)}\phi^{(1)} \nabla_{k}^{(0)}\phi^{(1)} + O\left(\frac{\epsilon^2}{\ell}\right).
\label{a.27+}
\end{eqnarray}

At last, let us consider the third term in the second line in (\ref{a.21+}). At the beginning we rewrite (\ref{b.5}) in the form:
\begin{eqnarray}
R^{(2)}_{km}= r_1+r_2+r_3+r_4,
\label{R1_4}
\end{eqnarray}
where
\begin{eqnarray}
r_1 &\equiv & -\frac{1}{2}\nabla_l^{(0)}\Big [ h^{lq} \big( \nabla_m^{(0)}h_{qk} + \nabla_k^{(0)}h_{qm} -\nabla_q^{(0)}h_{km} \big) \Big ],
\label{r1}\\
r_2 &\equiv & \frac{1}{2} \nabla_m^{(0)}\Big [ h^{lq} \nabla_k^{(0)} h_{ql}  \Big ], \label{r2}\\
r_3 &\equiv &\frac{1}{4} \nabla_n^{(0)} h \Big[ \nabla_m^{(0)} h^n_{k} + \nabla_k^{(0)} h^n_{m} -\nabla^{n\,(0)} h_{km} \Big],
\label{r3}\\
r_4 &\equiv &-\frac{1}{4}\Big[ \nabla_m^{(0)} h^l_{n} + \nabla_n^{(0)} h^l_{m} -\nabla^{l\,(0)}h_{nm} \Big]\Big[ \nabla_l^{(0)} h^n_{k} + \nabla_k^{(0)} h^n_{l} -\nabla^{n\,(0)}h_{kl} \Big].
\label{r4}
\end{eqnarray}

Now, substituting (\ref{a.14+}) into each of items (\ref{r1}) - (\ref{r4}) and realizing all the steps of the averaging procedure one obtains
\begin{eqnarray}
\langle r_1[\bar h_{ik}]\rangle &=& O\left(\frac{\epsilon^2}{\ell}\right),
\label{r1+}\\
\langle r_2[\bar h_{ik}]\rangle &=& O\left(\frac{\epsilon^2}{\ell}\right),
\label{r2+}\\
\langle r_3[\bar h_{ik}]\rangle &=& \frac{2}{f'^2_0} \nabla_k^{(0)}\phi^{(1)} \nabla_m^{(0)}\phi^{(1)} + O\left(\frac{\epsilon^2}{\ell}\right),
\label{r3+}\\
\langle r_4[\bar h_{ik}]\rangle &=& -\frac{1}{4}  \nabla_{k}^{(0)}\bar{h}_{nm}\nabla_{l}^{(0)}\bar{h}^{nm} -   \frac{3}{2}\frac{1}{f'^2_0} \nabla_k^{(0)}\phi^{(1)} \nabla_m^{(0)}\phi^{(1)} + O\left(\frac{\epsilon^2}{\ell}\right).
\label{r4+}
\end{eqnarray}
As a result, for averaged (\ref{R1_4}) one can derive
\begin{eqnarray}
\Big\langle f'_0  R^{(2)}_{ik}[\bar{h}_{ik}] \Big\rangle =   -\frac{f'_0}{4}\nabla_i^{(0)}\bar{h}^{nm}\nabla_k^{(0)}\bar{h}_{nm} +\frac{1}{2} \frac{1}{f'_0} \nabla_i^{(0)}\phi^{(1)}\nabla_k^{(0)}\phi^{(1)}  + O\left(\frac{\epsilon^2}{\ell}\right).
 \label{a.31+}
\end{eqnarray}

Finally, summing (\ref{a.24})-(\ref{phi2.average}), (\ref{a.25+}), (\ref{a.27+}) and (\ref{a.31+}), one obtains for the averaged (\ref{a.21+}):
\begin{eqnarray}
t_{ik}^{(dS)}
&=&
\frac{1}{32\pi G}
\left(  {f'_0}\nabla_i^{(0)}\bar{h}^{nm}\nabla_k^{(0)}\bar{h}_{nm} +6\frac{1}{f'_0} \nabla_i^{(0)} \phi^{(1)}  \nabla_k^{(0)}\phi^{(1)} \right) + O\left(\frac{\epsilon^2}{\ell}\right)
\label{a.33++C} \\
&=&
\frac{1}{32\pi G}
\left({f'_0}\partial_i\bar{h}^{nm}\partial_k\bar{h}_{nm} +6 \frac{1}{f'_0}\partial_i \phi^{(1)}  \partial_k\phi^{(1)} \right) + O\left(\frac{\epsilon^2}{\ell}\right),
\label{a.33}
\end{eqnarray}
where the definition $t_{ik}^{(dS)} \equiv - ({8\pi G})^{-1} \langle{J}_{ik}^{(2)}\rangle$ for the energy-momentum tensor  on the dS background has been introduced.
With making the use of (\ref{a.14+}), one has
\begin{eqnarray}
t_{ik}^{(dS)}
&= &
\frac{f'_0}{32\pi G}
\left( \nabla_i^{(0)}{h}^{nm}\nabla_k^{(0)}{h}_{nm} +\frac{1}{8}\nabla_i^{(0)} h  \nabla_k^{(0)}h\right)  + O\left(\frac{\epsilon^2}{\ell}\right)
\label{a.33+++С}\\
&=& \frac{f'_0}{32\pi G}
\left(\partial_ih^{nm}\partial_kh_{nm} +\frac{1}{8}  \partial_i h  \partial_k h  \right)+ O\left(\frac{\epsilon^2}{\ell}\right).
\label{a.33+}
\end{eqnarray}
In the result, we give concrete expressions (\ref{a.33}) and  (\ref{a.33+}) presented in the sets of variables $\{\bar h_{ik}, \phi^{(1)} \}$ and $\{h_{ik}, h \}$, respectively.

Concluding the section we remark that the energy-momentum of the gravitational waves is splitting onto three orders on the dS background
\begin{eqnarray}
t_{ik}^{(dS)} \rightarrow  O\left(\epsilon^2\right)  + O\left( \epsilon^2/\ell\right) +O\left( \epsilon^2/\ell^2\right),
\label{splitting+dS}
\end{eqnarray}
The leading order in both the covariant expressions (\ref{a.33++C}) and  (\ref{a.33+++С}) is defined by $O(\epsilon^2)$ that surpasses the next order $O\left({\epsilon^2}/{\ell}\right)$ because $\ell \gg 1$.  Thus, both non-covariant expressions (\ref{a.33}) and  (\ref{a.33+})) present the leading order purely. Such a situation is analogous to that in  \cite{A2}, see formula (4.1), where a covariant expression is splitting onto two orders $O(1)$ and $O(\epsilon)$ in notations of \cite{A1,A2} under transformation to non-covariant form.
Physically, the expression (\ref{a.33}) looks more preferable because the set  $\{\bar h_{ik}, \phi^{(1)} \}$ presents decoupled variables $\bar h_{ik}$ and  $\phi^{(1)}$,  each of them satisfies its own equation, and for which gauge freedoms have been used already. At last, the expression  (\ref{a.33+}) almost coincides with that obtained in \cite{BG1} on the initial flat background. The difference is in the multiplier $f'_0$ that reflects a fact that our consideration starts from the dS background.

\section{A back-reaction for gravitational waves in $f(R)$ theory }
\label{sect.5}

It is a textbook assertion that gravitational waves in GR brings energy and other energetic characteristics \cite{LL,MTW}. A one of approaches to show this is as follows. One provides expansions of the Einstein equations up to a second order in perturbations,  replaces the second order terms to the right hand side and interprets them as an effective gravitational wave energy-momentum.
The latter changes a curvature of spacetime by a back-reaction.


\subsection{The Isaacson procedure }

Such a problem has been studied by Isaacson \cite{A1,A2} who, applying the Brill-Hartle \cite{BH1} averaging procedure, has suggested a so-called high frequency limit. Gravitational waves are described by perturbations $h_{ik}$ with the amplitude of the order $h_{ik} \sim \epsilon \ll 1$. The notion ``high frequency'' is a relative notion. He considers two scales, first, the wavelength of gravitational ripple $\lambda$. Second, the characteristic scale $L$ of
a spacetime, the curvature value of which is calculated with taking into account the back-reaction of the averaged energy-momentum.
Namely, such a spacetime is used as a background spacetime for perturbations in the Isaacson scheme. Below, index ``$B$'' is introduced to denote background quantities in the Isaacson procedure, unlike index ``(0)'' used for the initial (fixed)  background quantities considered above. In another word, index ``$B$'' will be used for the quantities related to the averaged background spacetime.
Thus, vacuum Einstein equations are derived by Isaacson  as
\begin{eqnarray}
G^B_{ik} + G^{(1)}_{ik} +G^{(2)}_{ik} = 0,
\label{GR.eq}
\end{eqnarray}
where $G^B_{ik}$ is the Einstein tensor for the averaged background with the metric $g^B_{ik}$; the total metric is thought as $g_{ik} = g^B_{ik} + h_{ik}$. The second term in (\ref{GR.eq})  is linear in perturbations $h_{ik}$ on the averaged background, it defines the equation
\begin{eqnarray}
G^{(1)}_{ik}  = 0
\label{GR.G1}
\end{eqnarray}
that is just the gravitational wave equation determining $h_{ik}$. The third term in (\ref{GR.eq}) is quadratic in perturbations on the averaged background. Finally, averaging (\ref{GR.eq}) gives
\begin{eqnarray}
G^B_{ik} = - \langle G^{(2)}_{ik}\rangle = 8\pi G t^B_{ik},
\label{GR.eq+}
\end{eqnarray}
where $t^B_{ik}$ is just the averaged energy-momentum tensor of the gravitational waves of high frequency. Equations (\ref{GR.G1}) and (\ref{GR.eq+}) have to be solved simultaneously. Assuming that the wavelength of the gravitational ripple is $\lambda$, the averaging procedure has to be provided in regions of the scale $S$ for which $S\gg\lambda$.

The equation (\ref{GR.eq+}) gives a possibility to connect parameters of the averaged background and of the gravitational wave as
\begin{eqnarray}
O\left(\frac{1}{L^2}\right) = O\left(\frac{\epsilon^2}{\lambda^2}\right).
\label{L.lambda}
\end{eqnarray}
Thus,  Isaacson's scheme leads to the relation $\lambda \ll L$ that clearly supports the notion ``high frequency''. Note that the scales $S$ and $L$ really are not connected from the start. For $S$ one has only $S\gg\lambda$, whereas $L$ is defined by (\ref{L.lambda}) connecting to the gravitational wave parameters. However,  the restriction $S\gg L$ is not possible. Indeed,  in this case regions with the scale $L$ looks as a ripple with respect to scale $S$ that contradicts to the Isaacson averaging procedure.

Because there are the relative scales Isaacson uses a possibility to concretize a one of the them. He chooses $L=O(1)$, then by (\ref{L.lambda}) it turns out $\lambda = O(\epsilon)$.   As a result, first, $G^B_{ik} = O(1)$, second,  the important expressions are split as $G^{(1)}_{ik} = O(\epsilon^{-1}) + O(1) + O(\epsilon)$ and $t^B_{ik} = O(1) + O(\epsilon) + O(\epsilon^{2})$. Because the equation (\ref{GR.G1})  has to be hold in whole, the order $t^B_{ik} = O(1)$ defines the main value of averaged background uniquely. Indeed, the next order $t^B_{ik} = O(\epsilon)$ can be disturbed by the third order in expansion (\ref{GR.eq}), $G^{(3)}_{ik}$, not remarked there.  The Isaacson ideas initiated further development of his method, which has been explicitly applied to particular spacetimes in GR (see, for example, \cite{MCT1,Ta1,El1,Ar1,Ar2,Bu1,HF1,E1,An1,PS1,SP1}).

\subsection{The back-reaction on the dS background in $f(R)$ theory}
\label{back-reaction}

The Isaacson procedure has perspectives to be generalized to the equations in an arbitrary metric theory, and we do it for the case of the $f(R)$ theory with the dS space as an initial background. Following the works \cite{A1,A2,BG1} we use the Brill-Hartle average method \cite{BH1}.

Let us systemize parameters of the model. Following Isaacson, for gravitational ripple we define the amplitude of the order $\epsilon \ll 1$ and the wavelength $\lambda$. Thus, for the perturbations we have
\begin{eqnarray}
h_{ik} = O(\epsilon);\qquad\partial_n h_{ik} = O\left( \epsilon/\lambda\right);\qquad \partial_{mn}  h_{ik}= O\left( \epsilon/\lambda^2\right); \ldots ,
\label{h.lambda}\\
\phi^{(1)} = O(\epsilon);\qquad\partial_n \phi^{(1)}  = O\left( \epsilon/\lambda \right);\qquad \partial_{mn} \phi^{(1)} = O\left( \epsilon/\lambda^2\right); \ldots .
\label{phi.lambda}
\end{eqnarray}

To obtain the effective energy-momentum of gravitational waves we consider expansions of the quantity $J_{ik}$  obtained by variation with respect to $g^{ik}$ (the same as for
matter energy-momentum) and presented in (\ref{a.1}).
Now, analogously to (\ref{GR.eq}) we expand the equation (\ref{a.1}) on the averaged background spacetime (created with taking into account a back-reaction) with the metric $g^{B}_{ik}$ and scalar field $\phi^B$:
\begin{eqnarray}
J_{ik} &=& J^{B}_{ik} +  J^{(1)}_{ik} + J^{(2)}_{ik} = 0.
\label{J.B.1.2}
\end{eqnarray}
The quantity $J^{B}_{ik}$  is  defined by the expression in the left hand side of the  equation (\ref{a.1_0}) with changing index ``${(0)}$'' by index ``${B}$''.
The expression $J^{(1)}_{ik}$  is defined in (\ref{a.1_1}) for perturbations $h_{ik}$ and $\phi^{(1)}$ derived on an arbitrary background. Now it is defined by $g^{B}_{ik}$ and $\phi^B$, therefore, in the equation (\ref{a.1_1}), index ``${(0)}$'' has to be changed by index ``${B}$''. Thus, analogously to (\ref{GR.G1}) we have the linear equation
\begin{eqnarray}
J^{(1)}_{ik}  = 0.
\label{f(R).J1}
\end{eqnarray}
Now, let us average (\ref{J.B.1.2}) on the scale $S$ that is significantly more than the wavelength, but it is evidently that it cannot exceed the radius of cosmological horizon, thus  $S := \lambda\ll S \leq \ell$:
\begin{eqnarray}
 J^{B}_{ik} =-\langle J^{(2)}_{ik} \rangle & =& 8\pi G t^{B}_{ik}.
\label{J.1.2.angle}
\end{eqnarray}
Tensor $t^B_{ik}$ is the averaged energy-momentum tensor of the gravitational waves of high frequency, and $g^{B}_{ik}$ and $\phi^{B}$ satisfy the non-vacuum equation (\ref{J.1.2.angle}). Besides, equations (\ref{f(R).J1}) and (\ref{J.1.2.angle}) have to be solved simultaneously.

Due to $S \leq \ell$ parameters and expressions related to the initial de Sitter background are not subjected to changes under the averaging. Therefore, the equation (\ref{a.1_0}), $J^{(0)}_{ik}=0$, is preserved that means that the expression $ J^{(0)}_{ik}$ itself is not included into the left hand side of (\ref{J.1.2.angle}). Thus, $ J^{B}_{ik}$ is considered as induced by a back-reaction only. The quantities $g^{B}_{ik}$ and $\phi^{B}$ have to be thought as consisting of two parts:
\begin{eqnarray}
g^{B}_{ik}= g^{(0)}_{ik}+\tilde g^{B}_{ik}, \qquad \phi^{B}=\phi^{(0)}+\tilde\phi^{B},
\label{hB.phiB}
\end{eqnarray}
where $g^{(0)}$ and $\phi^{(0)}$ are the dS quantities, whereas $\tilde g^{B}_{ik}$ and $\tilde\phi^{B}$ can be interpreted as their shifts, respectively.

We specify $\tilde g^{B}_{ik}$ and $\tilde \phi^{B}$ by the characteristic length $\cal L$. Thus, $\tilde g^{B}_{ik}= O(1)$, $\partial_m \tilde g^{B}_{ik}= O(1/{\cal L})$, etc. and $\tilde \phi^{B}= O(1)$, $\partial_m \tilde \phi^{B}= O(1/{\cal L})$, etc. Then instead of (\ref{g0_sim+}) and (\ref{phi0_sim+}) one has the behaviour
%
\begin{eqnarray}
g_{ik}^{B}& =& O(1);\qquad\partial_n g_{ik}^{B} = O\left( 1/\ell\right) +O\left( 1/{\cal L}\right);\qquad \partial_{m}\partial_{n}  g_{ik}^{B}= O\left( 1/\ell^2\right)+O\left( 1/{\cal L}^2\right); \ldots ,
\label{g0_sim+B}\\
\phi^{B} &=& O(1);\qquad\partial_n \phi^{B} =O\left( 1/\ell\right) +O\left( 1/{\cal L}\right);\qquad \partial_{m}\partial_{n} \phi^{B}= O\left( 1/\ell^2\right)+O\left( 1/{\cal L}^2\right);  \ldots .
\label{phi0_sim+B}
\end{eqnarray}
Analogously, instead of (\ref{Gamma0_ell}) one has after applying (\ref{g0_sim+B}) for the behaviour of Christoffel symbols:
 \begin{eqnarray}
 \Gamma^{i\,B}_{kl} = O\left( 1/\ell\right)+O\left( 1/{\cal L}\right);\qquad\partial_n \Gamma^{i\,B}_{kl} = O\left( 1/\ell^2\right)+O\left( 1/(\ell{\cal L})\right)+O\left( 1/{\cal L}^2\right); \ldots.
\label{Gamma0_ellB}
\end{eqnarray}
After all the introduced above conventions one easily states from (\ref{J.1.2.angle}) the relation between the parameters of the model that generalized
(\ref{L.lambda}) in GR, it is
\begin{eqnarray}
O\left(\frac{ 1}{\ell{\cal L}}\right)+O\left(\frac{1}{{\cal L}^2}\right) = O\left(\frac{\epsilon^2}{\lambda^2}\right).
\label{L.lambda+}
\end{eqnarray}
The order $O({1}/{{\ell}^2})$ is not included because it is devoted to $J^{(0)}_{ik} = 0$.

 Because the scales $\lambda$ and ${\cal L}$ are relative ones we have a possibility to choose one of them as a unit. Unlike \cite{A1,A2},
 we choose $\lambda \sim O(1)$ that is more preferable by the next reasons. First, considering (\ref{h.lambda}) and (\ref{phi.lambda}),
 the choice $\lambda \sim O(1)$ preserves the behaviour in (\ref{h_sim}) and (\ref{phi_sim}) that makes the presentation in section \ref{sect.4}
 significantly simpler. Second, we have an external scale $\ell$ that  can be compared explicitly with ${\cal L}$.  (However, to have imaginations on all the scales in whole we will consider $\lambda$ evidently as well.)
 Thus, linear equations (\ref{a.1_1}) and (\ref{a.2_1}) are splitting to

\begin{eqnarray}
\rightarrow \qquad O\left(\epsilon\right) +  O\left({ \epsilon}/{\ell}\right)+ O\left({ \epsilon}/{\cal L}\right)+O\left({ \epsilon}/{\ell}^2\right)+O\left({ \epsilon}/({\ell}{\cal L})\right)+ O\left({ \epsilon}/{\cal L}^2\right)
\label{splitting+}
\end{eqnarray}
generalizing the splitting for linear equations (\ref{splitting}) on the purely dS background.
Concerning (\ref{L.lambda+}), it is interesting to consider three cases, 1) ${\cal L}\gg \ell$, 2) ${\cal L}\sim \ell$ and 3) ${\cal L}\ll \ell$.

In the case 1), the first term in (\ref{L.lambda+}) is of the leading order. Thus, it defines
\begin{eqnarray}
\epsilon^2\ell{\cal L} \sim \lambda^2 \sim 1.
\label{cal.L.ell}
\end{eqnarray}
The condition ${\cal L}\gg \ell$ means that a curvature induced by a back-reaction is very small. Then it is interesting to find out restriction when a back-reaction is negligible with respect to the dS background.
 It is achieved when all the additional terms in (\ref{splitting+}) are negligible with respect to all terms in splitting (\ref{splitting}).
Thus, it is necessary to strengthen the condition ${\cal L}\gg \ell$ by  $\lambda{\cal L}={\cal L}\gg \ell^2$ that satisfies this requirement. The latter together with  (\ref{cal.L.ell}) gives a restriction onto parameters of the gravitational wave $\epsilon^2 \ll \lambda^3/\ell^3 \sim 1/\ell^3$. The restriction  ${\cal L}\gg \ell^2$ makes negligible a back-reaction on the result of sections \ref{sect.4} with the splitting of the linear equations (\ref{splitting}) and of section \ref{AppendixC} with the splitting of the energy-momentum (\ref{splitting+dS}).

The cases 2) and 3) can be united into a one case because in (\ref{L.lambda+}) the first term becomes comparable or negligible with respect to the second one, and it gives the relation analogous to Isaacson's one (\ref{L.lambda}) that can be reformulated as
\begin{eqnarray}
\epsilon{\cal L} \sim \lambda \sim 1.
\label{cal.L.ell+}
\end{eqnarray}
Then, all the terms with $\ell$ in (\ref{splitting+}) are suppressed by the terms with  $\cal L$. Linear equations (\ref{a.1_1}) and (\ref{a.2_1}) are splitting to the main orders
\begin{eqnarray}
\rightarrow \qquad O\left(\epsilon\right) +   O\left({ \epsilon}/{\cal L}\right) + O\left({ \epsilon}/{\cal L}^2\right)
\label{splitting+L}
\end{eqnarray}
instead of (\ref{splitting}); and the energy-momentum has to be split to leading orders
\begin{eqnarray}
t_{ik}^{B} \rightarrow  O\left(\epsilon^2\right)  +  O\left({ \epsilon^2}/{\cal L}\right) + O\left({ \epsilon^2}/{\cal L}^2\right)
\label{splitting+B}
\end{eqnarray}
instead of (\ref{splitting+dS}). 

Let us compare the splitting of linear equations on the dS background (\ref{splitting}) and of linear equations on the averaged background (\ref{splitting+L}); the same, for the energy-momentum on the dS background (\ref{splitting+dS}) and on the averaged background (\ref{splitting+B}). One recognizes that they coincide in the leading terms without participation of $\ell$ or $\cal L$, for which $\ell\gg 1$ and ${\cal L} \gg 1$. Indeed, appearance of $\ell$ is induced by the covariant derivatives of the dS background, whereas a participation of $\cal L$ is induced by the covariant derivatives of the averaged background.
Then, providing the total program of sections \ref{sect.4} and \ref{AppendixC} but preserving the leading order only (preserving partial derivative only) one concludes that the leading orders both in the pair (\ref{splitting}) and (\ref{splitting+L}) and in the pair  (\ref{splitting+dS}) and (\ref{splitting+B}) coincide.

Thus, for the leading term in (\ref{J.1.2.angle}) of the order $O({\epsilon^2})$ one has
\begin{eqnarray}
t_{ik}^{B}
&= &
\frac{1}{32\pi G}
\left(f'_0\partial_i\bar{h}^{nm}\partial_k\bar{h}_{nm} +6 \frac{1}{f'_0}\partial_i \phi^{(1)}  \partial_k\phi^{(1)} \right)
\label{a.33Bbar}\\
&=& \frac{f'_0}{32\pi G}
\left(\partial_ih^{nm}\partial_kh_{nm} +\frac{1}{8}  \partial_i h  \partial_k h  \right),
\label{a.33B}
\end{eqnarray}
the same as in (\ref{a.33}) and  (\ref{a.33+}).

Analogously to the Isaacson picture, only the leading term in (\ref{splitting+B}), that is $O\left(\epsilon^2\right)$ concretely expressed in (\ref{a.33Bbar}) (the same in (\ref{a.33B})), plays a crucial role. Indeed, by (\ref{cal.L.ell+}) the term $J^{(3)}_{ik}$ (not included in (\ref{J.B.1.2})) disturbs the second term in  (\ref{splitting+B}).

\section{Concluding remarks and discussion}
\label{sect.6}


The main goal of the paper is a construction of the averaged energy-momentum for high frequency gravitational waves in $f(R)$ theory.

First, by the reasons given in Introduction, we carry out our study in the scalar-tensor form of $f(R)$ where  the scalar massive mode appears automatically as $\phi^{(1)}$. Such a possibility is used by other authors to study various problems in $f(R)$ theory as well, see, for example, \cite{1106.5582,1608.01764,1701.05998}.
However, we have pointed out the connection between   the scalar-tensor and the tensor-geometrical forms. A transition from one to another is given in subsection \ref{s-t-repres}. In the linear approximation, the connection of the scalar and tensor perturbations are given by (\ref{b.7a}) and by (\ref{b.7a++}). Redefinition of the scalar and tensor perturbations in (\ref{a.13}) and (\ref{a.14}) is necessary to decouple the related equations. All of these is permissible because both of the forms are equivalent. Finally, we present the averaged energy-momentum both in the scalar-tensor form (\ref{a.33Bbar}) and in the pure tensor form (\ref{a.33B}).

Second, a necessary step in the study is derivation  and analyse of linear equations. We have started from an arbitrary background. The gauge transformations and gauge invariance were discussed in detail and used maximally. The novelty is that a restriction to backgrounds presented by the Einstein spaces \cite{PetrovAZbook} allows us to decouple linear equations onto tensor and scalar parts. The restriction to dS background allows us easily to apply a so-called TT-gauge and derive the equations in the simplest way. In this case, the interesting result is that the linear tensor equations exactly coincide with those in GR with the Einstein cosmological constant on dS background. Thus, we generalize the claim in \cite{BG1} obtained on a flat background in $f(R)$ theory.

Third, we construct the averaged energy-momentum on the dS background, see (\ref{a.33}) and (\ref{a.33+}). It is quite instructive  because it allows us to provide all the calculations completely in spite of they are very complicated.

Fourth, following the Isaacson procedure we analyze influence of a back-reaction. The first case is presented by a situation when the back-reaction is very weak and the results in sections \ref{sect.4} and \ref{AppendixC} (on the dS background) are left unchanged. It is achieved when the characteristic length of the background $\cal L$ induced by a back-reaction is very big (consequently, the related curvature is very small) and it is estimated as ${\cal L}\gg \ell^2/\lambda$. Then parameters of the gravitational wave have to be restricted by $\epsilon^2\ll \lambda^3/\ell^3$. For the second case, when $\cal L$ is comparable with the dS radius or less it, the scheme of the sections \ref{sect.4} and \ref{AppendixC} is disturbed.  It takes a place when ${\cal L} \sim \lambda/\epsilon $ analogously to the Isaacson relation (\ref{L.lambda}). In the result we obtain for the averaged energy-momentum on the curved by a back-reaction background the result (\ref{a.33Bbar}) and (\ref{a.33B}) that coincides with (\ref{a.33}) and (\ref{a.33+}) in the leading order.

The notion of ``high frequency'' takes a place with respect to all of the introduced scales. Thus, one has simultaneously
\begin{itemize}

\item[1)] $\lambda \ll S$, where the scale $S$ was introduced for the average procedure; because $\ell$ is the cosmological horizon radius $S\leq\ell$, and $S$ is independent on $\cal L$,

\item[2)] $\lambda \ll \ell$ because $\ell$ is the cosmological horizon radius,

\item[2)] $\lambda \ll {\cal L}$ due to the construction of section \ref{sect.5}.

\end{itemize}

In spite of (\ref{a.33Bbar}) and (\ref{a.33B}) almost repeat the result of \cite{BG1} with an initial flat background,  there is a difference in multiplier $f'_0$. This just signals that for the starting background the dS space was chosen. Indeed, although $f'_0 = O(1)$ it can be differed from the unit, for example, for the theory (\ref{Exp.L}) one has $f'_0 = 1-\beta e^{-1}$.

Of course, there are many modifications, developments and applications of the Isaacson scheme. For example, in the papers \cite{E1} and \cite{SP1}, the authors study in GR the relation between  $\lambda$ and $L$ with the goal to find restrictions on the high frequency limit. In the paper \cite{KAM1}, the authors suggest a formulation for
corrections of the geometrical optics expansion in the framework of the Horndeski theory that includes five parameters for describing the amplitude and wavelength of the gravitational ripple, and different scales of other characteristic lengths. In a certain sense, we develop the Isaacson procedure by an analogical way considering parameters $\epsilon$, $\lambda$, $\ell$, $\cal L$ and $S$.

Finally, let us discuss the behaviour (\ref{prime}) that ensures the results (\ref{a.33}), (\ref{a.33+}) and (\ref{a.33Bbar}), (\ref{a.33B}) of sections \ref{AppendixC} and \ref{sect.5}, respectively. Really, these requirements could be weakened, for example, as
\begin{eqnarray}
\frac{f'_0}{f''_0} \sim  \frac{f''_0}{f'''_0}\sim O\left(\frac{1}{\ell}\right),
\label{prime+}\\
\frac{f'_0}{f''_0} \sim  \frac{f''_0}{f'''_0}\sim O\left(\frac{1}{{\cal L}}\right)
\label{primeB+}
\end{eqnarray}
instead of (\ref{prime}). Then one needs to find out a related restrictions for functions $f(R)$. More weak restrictions than (\ref{prime+}) and (\ref{primeB+}) could change the results (\ref{a.33}), (\ref{a.33+}) and (\ref{a.33Bbar}), (\ref{a.33B}) for the averaged energy-momentum in $f(R)$ theory, that looks interesting. All of these is a matter for our further studies, and we plan this.

\begin{acknowledgments}
PT thanks S. Sushkov for useful discussion; AP: the study was conducted under the state assignment of Lomonosov Moscow State University; the authors thanks the unknown referees for the useful comments and fruitful recommendations.
\end{acknowledgments}

\appendix

\section{Behaviour of  $f'_0/f''_0$ and $f''_0/f'''_0$  on the dS backgrounds }
\label{AppendixA+}

To analyze the expressions (\ref{a.16.1+}) and (\ref{a.21+}) it was used the behaviour
\begin{eqnarray}
\frac{f'_0}{f''_0} \sim  \frac{f''_0}{f'''_0}\sim {R^{(0)}}= O\left(\frac{1}{\ell^2}\right).
\label{prime}
\end{eqnarray}
By this the related terms in (\ref{a.16.1+}) and (\ref{a.21+}) become negligible and one achieves the final result in (\ref{a.33}) and  (\ref{a.33+}). Once again emphasizing the importance of
the requirements (\ref{prime}) and following them, we declare the main goal of this Appendix as a demonstration of the fact that Lagrangians satisfying (\ref{prime}) exist.

Besides of (\ref{prime}), it is important to note that the desired Lagrangians have to permit a late
time acceleration phase described by dS space, on the background of which gravitational waves are considered here.
From the first glance one could follow \cite{BG1} to represent $f(R)$ as a power series
\begin{eqnarray}
f(R) = R + \frac{a_2}{2!}R^2 + \frac{a_3}{3!}R^3 + \dots\, .
\label{series}
\end{eqnarray}
One can show that functions, like (\ref{series}), satisfy (\ref{prime}). However, first, such a form was suggested in \cite{BG1} for the case of Minkowski background, whereas to describe dS expansion the expression (\ref{series}) has to contain a constant term. Second, a power series with only the positive degrees can describe only the inflation stage and cannot describe the stage of the modern acceleration \cite{SF1}, like $f(R) = R + \alpha R^2$ in the pioneer work by Starobinsky \cite{Starobinsky.0}.

Therefore it is more preferable to turn to the representation (\ref{fR.series}) that can be rewritten in the form
\begin{eqnarray}
f(R) = {\sum}^{+\infty}_{-\infty} a_n (R - R_n)^n\,,
\label{seriesN}
\end{eqnarray}
where the present time acceleration stage could be modeled by inverse power law terms. Of course, it is assumed that the power series (\ref{seriesN}) converges.
Then, returning to the equation (\ref{1.2}), if one finds physically relevant solutions, one has to conclude that related  power series $f'(R),~f''(R),~f'''(R)$ have to converge as well. Finally, because $f'(R),~f''(R),~f'''(R)$ consists of power terms they have to satisfy (\ref{prime}).

As an example of an exact function without expansions we derive the Lagrangian of
a so-called Exponential Gravity \cite{0712.4017,1012.2280,Linder}:
\begin{eqnarray}
f(R) = R - \beta R_S\left(1- e^{-R/R_S} \right),
\label{Exp.L}
\end{eqnarray}
where $\beta$ and $R_S$ are constants. Setting $R_S = R^{(0)}$ one easily checks that the behaviour  (\ref{prime}) holds.

Remark, the behaviour (\ref{prime})  makes a behaviour of the relative terms in  (\ref{a.16.1+}) and (\ref{a.21+}) negligible with respect to a leading orders. However, (\ref{prime}) are very strong. They can be done more weak to eliminate the problem of disturbing the leading orders.

\end{document}